\documentclass[useAMS,usenatbib]{mn2e}
\usepackage{epsfig}
\usepackage{amsmath}
\usepackage{graphicx}
\usepackage{pdfpages}
\title[SFR in 2MIG isolated galaxies]
  {Star formation rates in isolated galaxies selected from the Two-Micron All-Sky Survey}

  \author[O. Melnyk et al.]
  { O.~Melnyk$^{1,2}$, V. Karachentseva$^{3}$, I. Karachentsev$^{4}$\\
$^1$Astronomical Observatory, National Taras Schevchenko University of Kyiv, 3 Observatorna St., 04053 Kyiv, Ukraine\\
$^2$Dipartimento di Fisica e Astronomia, Universit\`a di Bologna, Viale Berti Pichat 6/2, I-40127  Bologna, Italy \\
$^3$Main Astronomical Observatory, Academy of Sciences of Ukraine, 27 Akademika Zabolotnoho St., 03680 Kyiv, Ukraine\\
$^4$ Special Astrophysical Observatory of the Russian Academy of Sciences, Nizhnij Arkhyz, KChR, 369167, Russia\\
}
\date{Released 2013 Xxxxx XX}

\pagerange{\pageref{firstpage}--\pageref{lastpage}} \pubyear{2013}

\def\LaTeX{L\kern-.36em\raise.3ex\hbox{a}\kern-.15em
    T\kern-.1667em\lower.7ex\hbox{E}\kern-.125emX}

\begin{document}

\label{firstpage}

\maketitle

\begin{abstract}
We have considered the star formation properties of 1616 isolated galaxies from the 2MASS XSC selected sample (2MIG)
with the FUV GALEX magnitudes. This sample was then compared with corresponding properties of isolated galaxies from the Local 
Orphan Galaxies catalogue (LOG) and paired galaxies. 

We found that different selection algorithms define different populations of isolated galaxies. The population of the
LOG catalogue, selected from non-clustered galaxies in the Local Supercluster volume, 
mostly consists of low-mass spiral and late type galaxies. The $SSFR$ upper limit in isolated and paired galaxies 
does not exceed the value of $\sim$dex(-9.4). This is probably common for galaxies
of differing activity and environment (at least at z$<$0.06).
The fractions of quenched galaxies are nearly twice as high 
in the paired galaxy sample as in the 2MIG isolated galaxy sample.
From the behaviour of $(S)SFR$ vs. $M_*$ relations we deduced that the characteristic value influencing evolutionary processes 
is the galaxy mass. However the environmental influence is notable: paired  massive galaxies with log$M_*>$11.5 have 
higher $(S)SFR$ than isolated galaxies. Our results suggest that the environment helps to trigger the star 
formation in the highest mass galaxies. We found that the fraction of AGN in the
paired sample is only a little higher than in our isolated galaxy sample. 
We assume that AGN phenomenon is probably defined by secular galaxy evolution. 

\end{abstract}


\section{Introduction}
Recent observational and theoretical results have shown that a peak of the star formation rate ($SFR$) appears between z
$\sim$2-3, decreasing to the present epoch (Bouwens et al. 2011, Burgarella et al. 2013, Behroozi et al. 2013 and references therein).
This is likely due to the fact that evolution of bimodality in colour distribution is clearly observed, showing that the fraction of 
red massive (i.e. early type) galaxies with quenched star formation is higher at z$\sim$ 0 than in the earlier epochs. Star formation rate 
also depends on the environmental density, assuming that quenching of star formation is more efficient in high density regions 
(Cucciati et al. 2006, Scoville et al. 2013, Tal et al. 2014). 

Meanwhile, the dominant factor of quenching is different for the high mass (central) and low mass 
(satellite) galaxies. According to findings by Peng et al. (2010, 2012), the former galaxies preferred the self-regulated 
``mass quenching'', while quenching in latter galaxies occurred mainly due to environmental and/or merging influence 
(see also Tinker et al. 2013, Tal et al. 2014, Wetzel et al. 2014). 
However, the physical mechanisms of internal ``mass quenching'' are still under discussion. 
It would appear that active galactive nuclei (AGN) feedback is the dominant regime for internal mass quenching.
Occurring in high mass galaxies, it consists in kinematic outflows (winds) from the accretion disk surrounding the black hole. 
This removes the cold gas from the galaxies, terminating the star formation. 
The theoretical underpinnings of the importance of this process are described by Khalatyan et al. (2008), Dubois et al. (2013) 
and others. The observational evidence is summarized in Fabian (2012), see also a statistical study by Lemaux et al. (2013). 
It is also noted that radio feedback observed in radio-loud AGN plays an important role in the host galaxy star 
formation quenching, see for example Croton et al. (2006).  For example, Schawinski et al. (2014) argues in a that AGN feedback is a key 
factor accompanying spiral galaxy merging for quenched elliptical galaxy formation. Apparently this process could explain a
formation of isolated/void early type galaxies, where an influence of environment is minimal (Croton \& Farrar 2008).
It is thought that AGN feedback influence could also be positive, triggering star formation on shorter timescales 
(Zinn et al. 2013, Zubovas 2013). In addition to minor and major mergings, the processes of ram pressure stripping, harassment and strangulation 
facilitate environmental quenching. Relatively slow strangulation, where a galaxy loses its gas due to tidal effects produced 
by the gravitational potential of the group/cluster seems to be a more effective quenching mechanism than others 
(van den Bosch et al. 2008, Taranu et al. 2014). 

An effective method for the study of environmental impact upon galaxy evolution is to compare the properties of galaxies from dense environments
with corresponding properties of isolated galaxies, i.e. objects that have not been appreciably affected by an external influence 
of their environments for at least a few Gyr (see Verley et al. 2007, Karachentseva et al. 2010). 
This has been confirmed by comparison between modelling results and observations, revealing
that properties of isolated galaxies have been mainly influenced by internal processes. 
Hirschmann et al. (2013) showed that only 45\% of isolated galaxies have experienced at least one merger
event in the past (most of the mergers are minor, with mass ratios between 1:4 and 1:10). 
More results and discussions on the small satellites search and their influence on isolated galaxy properties
can be found in works by Verley et al. (2007), Karachentseva et al. (2011), Argudo-Fernandez et al. (2013) and Melnyk et al. (2014).  

We note that the isolated galaxies are objects 
selected in projection on the sky or in velocity (3D) space with basic criteria of "no significant companions" in 
a given volume (examples of the different approaches can be found in the works by Karachentseva 1973, Varela et al. 2004, Allam et al. 2005,  Elyiv et al. 2009, 
Karachentseva et al. 2010, Karachentseva et al. 2010a, Hernandez-Toledo et al. 2010, Karachentsev et al. 2011).

This is the main difference between isolated galaxies and ``non-clustered'' or
``field'' galaxies. These are objects located outside of groups and clusters (de 
Vaucouleurs 1971, Turner \& Gott 1977, Makarov \& Karachentsev 2011, Karachentsev et al. 2012), 
in addition to void galaxies, selected as objects 
residing in low density regions (Rojas et al. 2004, Sorrentino et al. 2006, Patiri et al. 2006, 
von Benda-Beckmann \& Muller 2008, Hoyle et al. 2012, Kreckel et al. 2012, Elyiv et al. 2013).

The most discussed sample of isolated galaxies is the Catalogue of Isolated Galaxies (Karachentseva 1973, KIG). \footnote{The 
catalogue is known as KIG in Russian transcription but the abbreviation CIG is also widely used.}
The catalogue includes 1050 galaxies with  $m \leq 15.7$ 
and $\delta>-3^{\circ}$, i.e. $\sim$4\% in the CGCG catalogue (Zwicky et al. 1961-1968). The largest contribution to the analysis of  
the KIG properties was done by the AMIGA team, with their most important results in Verley et al. (2007),
Lisenfeld et al. (2007, 2011), Durbala et al. (2008),  Leon et al. (2008), Sabater  et al. (2008, 2012), Fernandez Lorenzo et al. (2012) and Argudo-Fernandez et al. (2013).

In  Karachentseva et al. (2010) we applied slightly modified Karachentseva (1973) isolation criteria to the 
Two Micron All-Sky Survey Extended Source Catalog (2MASS XSC; Jarrett et al. 2000) to compile the sample of near-infrared isolated galaxies (2MIG;
see details in the next section). The properties of galaxies from the 2MIG catalogue have been investigated in several works.   
Karachentseva et al. (2011) estimated the orbital masses and mass-to-luminosity ratios of isolated galaxies with faint companions. It was 
shown that these companions have a weak influence on the dynamic isolation of the 2MIG galaxies. Kudrya et al. (2011)  considered the 
statistical relations between different observational characteristics while 
Kudrya \& Karachentseva (2012) constructed the Tully-Fisher relations for the 2MIG galaxies. Coziol et al. (2011) 
made an extensive analysis of AGN impact in the 2MIG sample, concluding that this AGN phenomenon is 
closely connected with the internal factor of galaxy evolution. Pulatova et al. (2015) reached a similar conclusion, making a multiwavelength analysis
of 36 2MIG galaxies displaying AGN. Anderson et al. (2013) investigated a large sample of 
X-ray selected 2MIGs observed with ROSAT, reporting an extended emission presence around a significant part of the isolated galaxy sample.
Melnyk et al. (2014) considered the near-infrared and optical colours of the 2MIG galaxies, comparing them with corresponding colours of 
galaxies located in denser regions. It was found that in general, 2MIG galaxies have a bluer colour than galaxies in pairs/groups, 
except the most compact pairs which are a little bluer than the 2MIGs, possibly due to recent merging events. 

In this paper we also consider the Local Orphan Galaxies catalogue (LOG; Karachentsev et al. 2011), which was compiled from the 
field (non-clustered) galaxies within the Local Supercluster volume by applying the 
isolation criteria used by Karachentseva (1973). Applying the KIG isolation criterion to the flux-limited CGCG galaxy catalogue, Karachentseva (1973)
selected $\sim$4\% isolated galaxies. The  use of the slightly modified KIG criterion to the 2MASS XSC catalogue (also flux-limited) giving $\sim$6\% isolated galaxies.
Essentially, we assume that isolated galaxies exist but are not numerous. When compiling the LOG catalogue (volume-limited) 
we took these percentages into account and fixed the 5\% level as typical for isolated galaxies.

Karachentsev et al. (2013) studied the star formation properties 
of the LOG galaxies and found that the specific star formation rate ($SSFR=SFR/M_*$) upper limit does not exceed the value of 
log$SSFR$ = -9.4 [$yr^{-1}$]. This agrees with the corresponding upper limits for the galaxies from the Local Universe 
located in different environments (Karachetsev \& Kaisina 2013) and for galaxies of different activity 
types in the Markarian sense (Karachentseva et al. 2014).
 
The current paper is a continuation of our previous investigations of basic properties of the 2MIG isolated galaxy sample, focusing 
upon the $SFR$ properties. We aim to develop a better understanding of the level of the star formation in the Local Universe. We define this as
$SFR$ and $SSFR$ vs. stellar mass $M_*$ and other relations for the 2MIGs and compare them with corresponding relations for the LOG isolated 
galaxy population, compiled from a different primary sample with different selection conditions. We also study the environmental influence
on star formation quenching/triggering, comparing the 2MIG properties with those of galaxies from the wide/compact pairs.

In Section 2 we present the isolation criteria for the 2MIG and LOG samples selection and describe data used in the paper. 
In Section 3 we discuss the star formation properties of the 2MIG and LOG isolated galaxies. 
The results of a comparison of the basic $SFR$ relations and AGN contamination for isolated and paired galaxies are given in Section 4. 
In Section 5 we give a summary of our results and the main conclusions.

\section {The data}

\subsection {The 2MIG catalogue}
 
The 2MIG entire sky catalogue of isolated galaxies (Karachentseva et al. 2010) was selected from 1.6 million objects of the 
2MASS XSC (Jarrett et al. 2000).  Selection of objects was performed twice: 

(i) automatically according to the original criteria of isolation by Karachentseva (1973) adapted to the 2MASS data:

\begin{eqnarray} 
X_{1i}/a_i\geq s = 30
\label{lpar:eq}
\end{eqnarray}

and 

\begin{eqnarray}
4\geq a_i/a_1 \geq 1/4, 
\label{lpar:eq}
\end{eqnarray}

where subscripts “1” and “i” refer to the fixed galaxy and its neighbours, respectively. Essentially, 
a galaxy with a standard angular diameter $a_1$ is considered to be isolated if its angular separation $X_{1i}$ from all its neighbours 
with “significant” angular diameters $a_i$ inside interval (2) is equal to or exceeds 30$a_{i}$.

(ii) visual inspecting of the images (DSS1,2\footnote{Digital Sky Survey: http://archive.eso.org/dss/dss} and 
SDSS\footnote{Sloan Digital Sky Survey: http://www.sdss.org})  of all isolated candidates to identify visible neighbours. 
Visual inspection allowed us to physically eliminate multiple systems with blue neighbours missed in the 
automatic selection (see details in Karachentseva et al. 2010).  Finally, the 2MIG catalogue consists of 3227 galaxies brighter 
than $K_s$ = 12 mag and with angular diameters $a_{Ks}\geq $ 30$''$.  We found approximately 6\% of isolated galaxies 
among the 51572 extended sources of the 2MASS XSC survey.

In total, 2869 from the 3227 2MIG galaxies have radial velocities according to the NED\footnote{NASA/IPAC Extragalactic Database: 
\\ http://ned.ipac.caltech.edu} and HyperLeda\footnote{Database for physics of galaxies: http://leda.univ-lyon1.fr/ (Paturel et al. 2003).} 
databases. We removed 23 galaxies  from the 2MIG list due to the presence of significant companions in their 
neighbourhood, or otherwise their protometry was contaminated by a projected star: 2MIG90, 174, 243, 320\footnote{In this work 
2MIG320=NGC1050 is still a part of our sample since we found a significant satellite when the text of the paper was finalized.}, 586, 1137, 1172, 1548, 1809, 1819, 1822, 1949,
1970, 1988, 2242, 2243, 2249, 2285, 2585, 2976, 2989, 3143, 3193, 3201. Therefore, in this paper we consider only 1616 isolated galaxies 
having FUV magnitudes (see next subsection). The morphological types of galaxies in the paper correspond to the digital scale: 
-2 -- E, 0 -- S0, 1 -- S0a/Sa, 2 -- Sab, 3 -- Sb, 4 -- Sbc, 5 -- Sc, 6 -- Scd, 7 -- Sd, 8 -- Sdm, 9 -- Im, 10 -- Ir. We note that 
near 7\% of the catalogue galaxies have different morphologies in comparison to the types noted in primary 2MIG catalogue 
(Karachentseva et al. 2010). These morphological types were revised 
according to SDSS DR10 images and the classification used by Coziol et al. (2011), see also Melnyk et al. (2014).

\begin{figure}
\includegraphics[width=0.45\textwidth,natwidth=500,natheight=400]{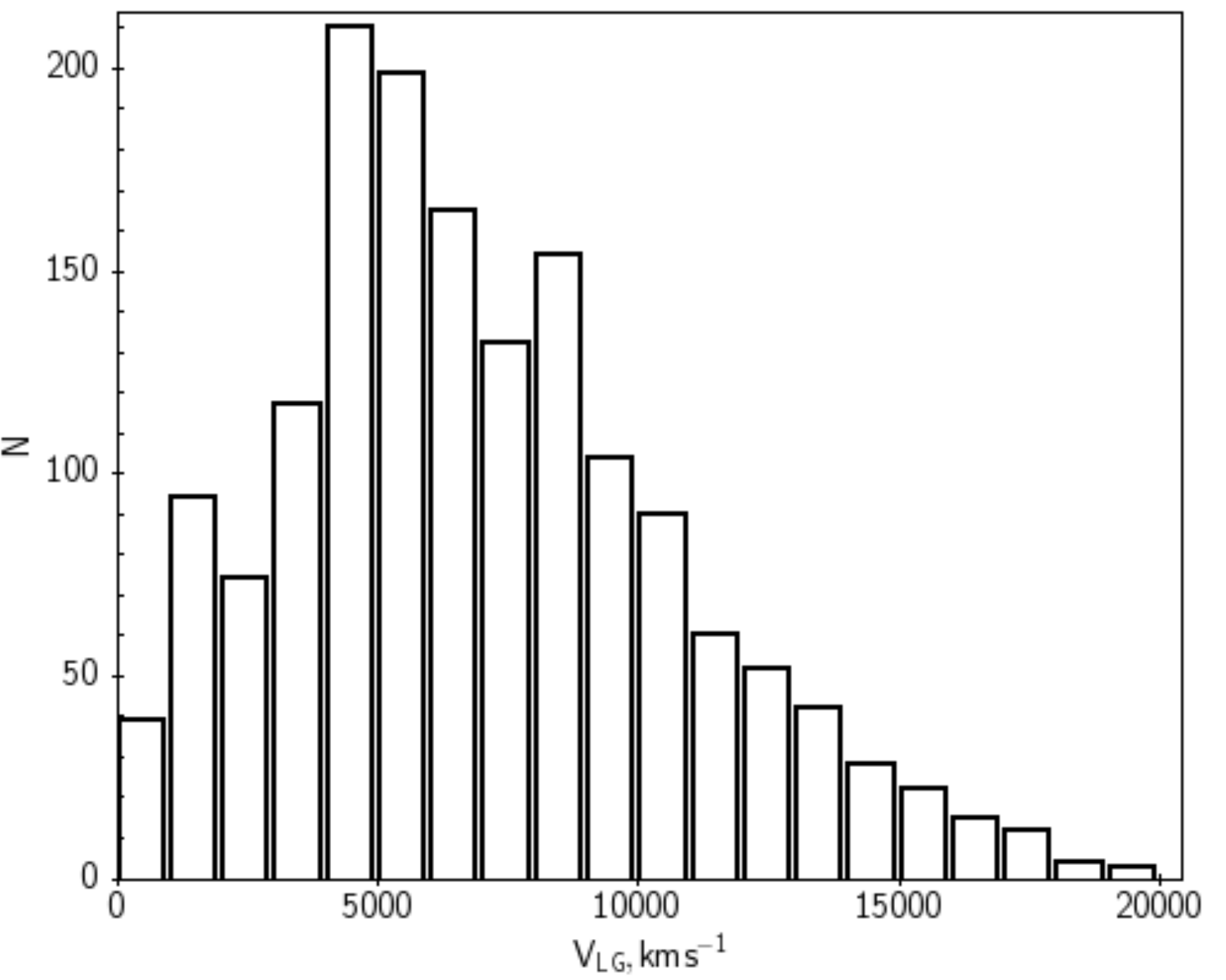}
\caption{Radial velocity distribution for the full 2MIG sample considered in this work, N=1616.}
\label{1}
\end{figure}

\subsection {The LOG catalogue}

The isolated galaxies for the LOG list were selected from the primary catalogue of $\sim$11000 galaxies within the volume of the Local 
Supercluster $V_{LG}<$3500 km/s at the Galactic latitudes $ \lvert b \rvert > 15^\circ$. At first, only galaxies from the primary galaxy catalogue of the virialized groups 
were selected. For that, Makarov \& Karachentsev (2011) used a modified percolation method. This takes into account the individual 
characteristics of galaxies and the following conditions: 

\begin{eqnarray}
\frac{V_{12}R_{12}}{2GM_{12}} < 1,
\end{eqnarray}

where $V_{12}$ is the velocity difference between two galaxies, $R_{12}$ is the corresponding projected distance, $G$ is the gravitational 
constant 
and $M_{12}$ is the total mass of the pair. Each galaxy mass was calculated using condition $M/L_K=\kappa M_{\odot}/L_{\odot}$, $L_K$ is 
the 2MASS  $Ks$ band luminosity while $\kappa$=6 is the empirical coefficient. The condition (3) means that a total energy of the physical pair must be 
 negative and supplemented by condition
(4), according to which the pair of components should remain within the sphere of 'zero-velocity':

\begin{eqnarray}
\frac{\pi H_0^2R_{12}^3}{8GM_{12}} < 1,
\end{eqnarray}

where $H_{0}$ is the Hubble constant.  All the pairs satisfying conditions (3) and (4) with a common main component were combined into a 
group. Therefore, a primary  sample of $\sim$11000 galaxies was divided  into ``clusterized'' galaxies, i.e. 54\% are located in virialized groups, 
and ``non-clusterized'' galaxies. The candidates for isolated galaxies were chosen by application of the clusterization algorithm to all 
galaxies with empirical coefficient $k=\kappa \cdot$ 40=240. If, at this strong condition, a galaxy still remained ``non-clustered'' it was 
considered as a candidate to the isolated one (see more details in Karachentsev et al. 2011). Among 990 isolated candidates, the true isolated 
galaxies were chosen by application of the Karachentseva (1973) selection criteria, i.e. visual inspection of all targets and their environments. 
As a result, the LOG catalogue consists of 520 isolated galaxies, 3 of which were excluded as non-isolated 
(Karachentsev et al. 2013). In this paper we consider 428 LOG galaxies with $FUV$ measurements (see subsection 2.3).

\begin{figure*}
\tabcolsep 0 pt
\begin{tabular}{ccc}
\includegraphics[width=0.35\textwidth,natwidth=500,natheight=400]{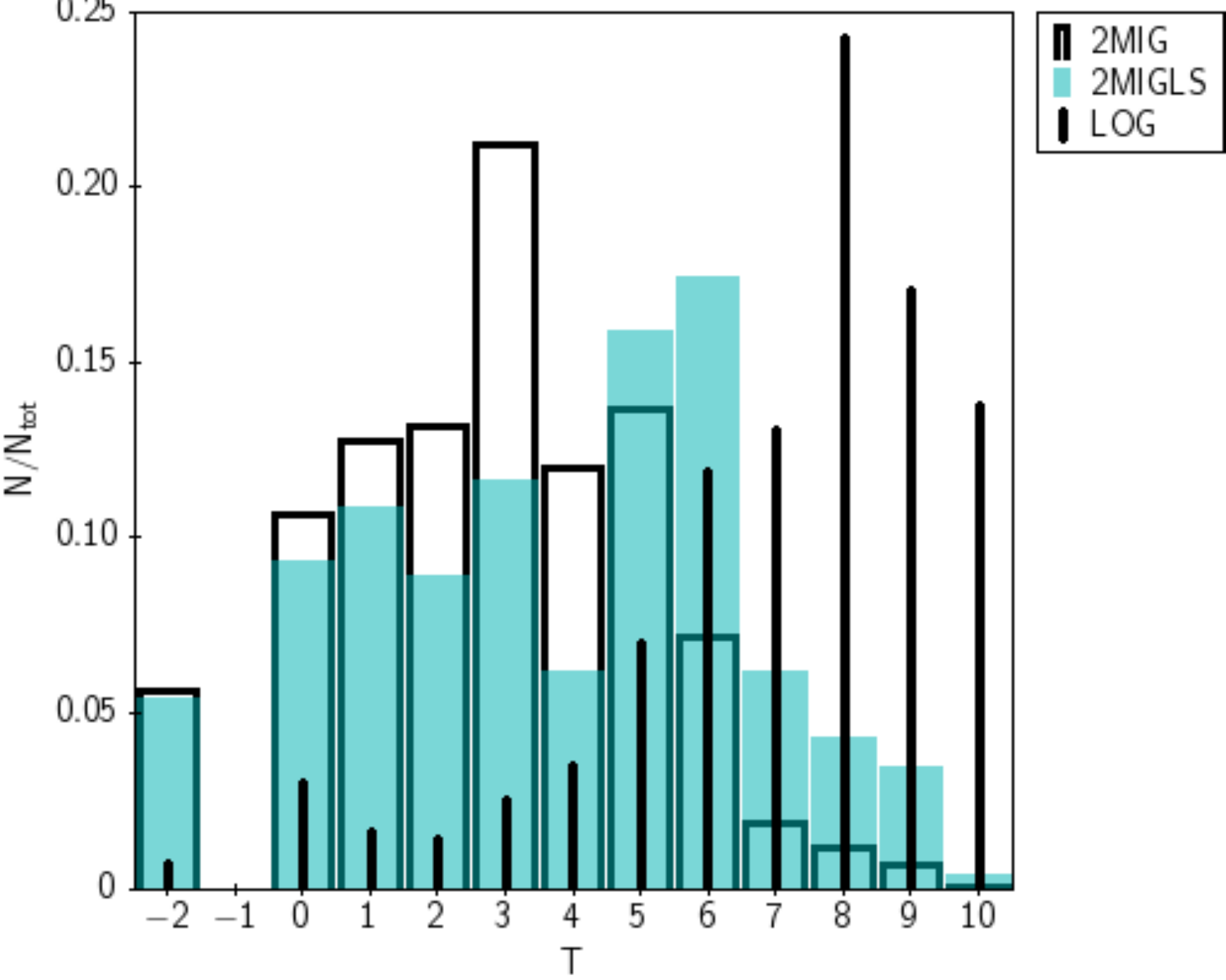}
\includegraphics[width=0.35\textwidth,natwidth=500,natheight=400]{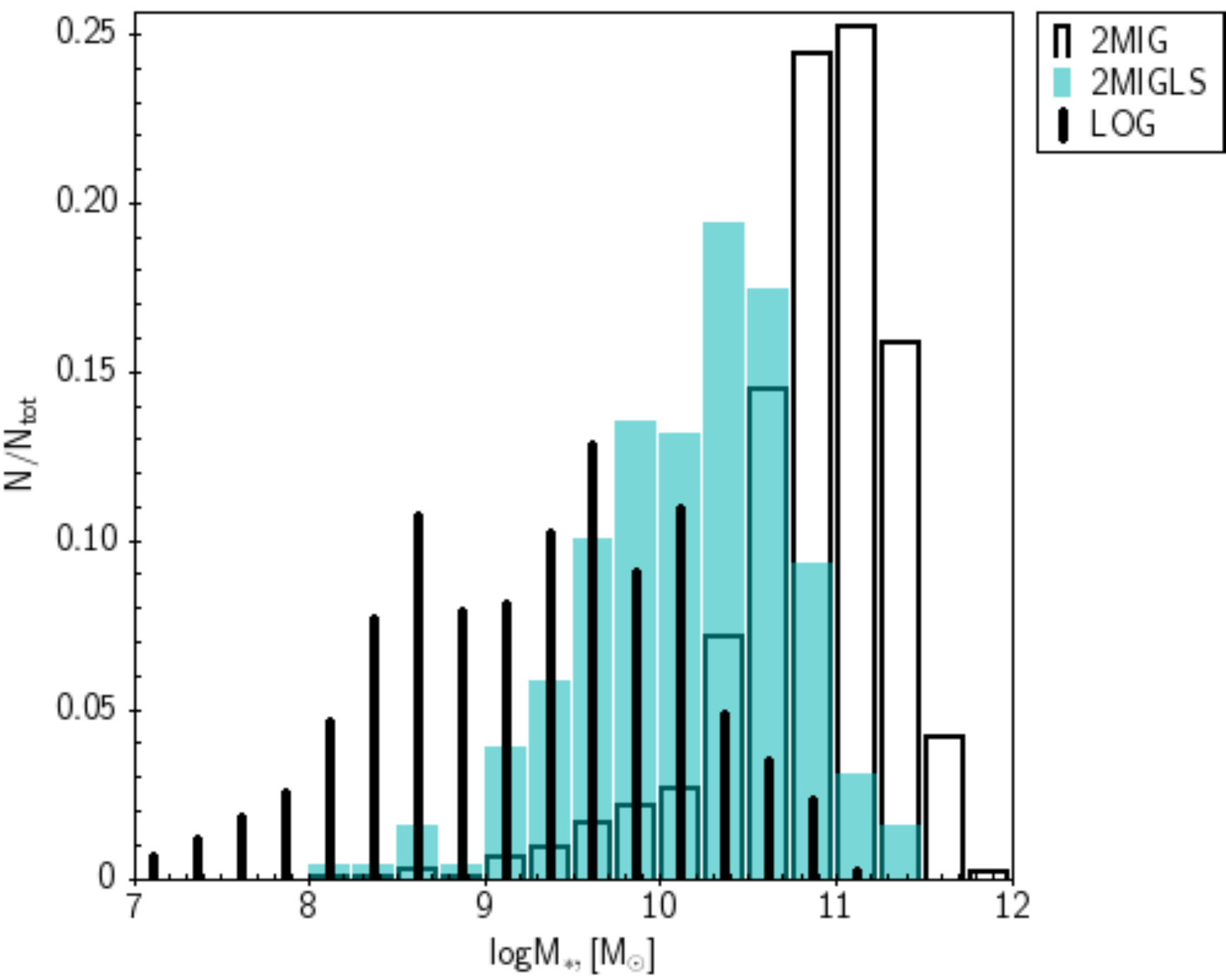} &
\includegraphics[width=0.35\textwidth,natwidth=500,natheight=400]{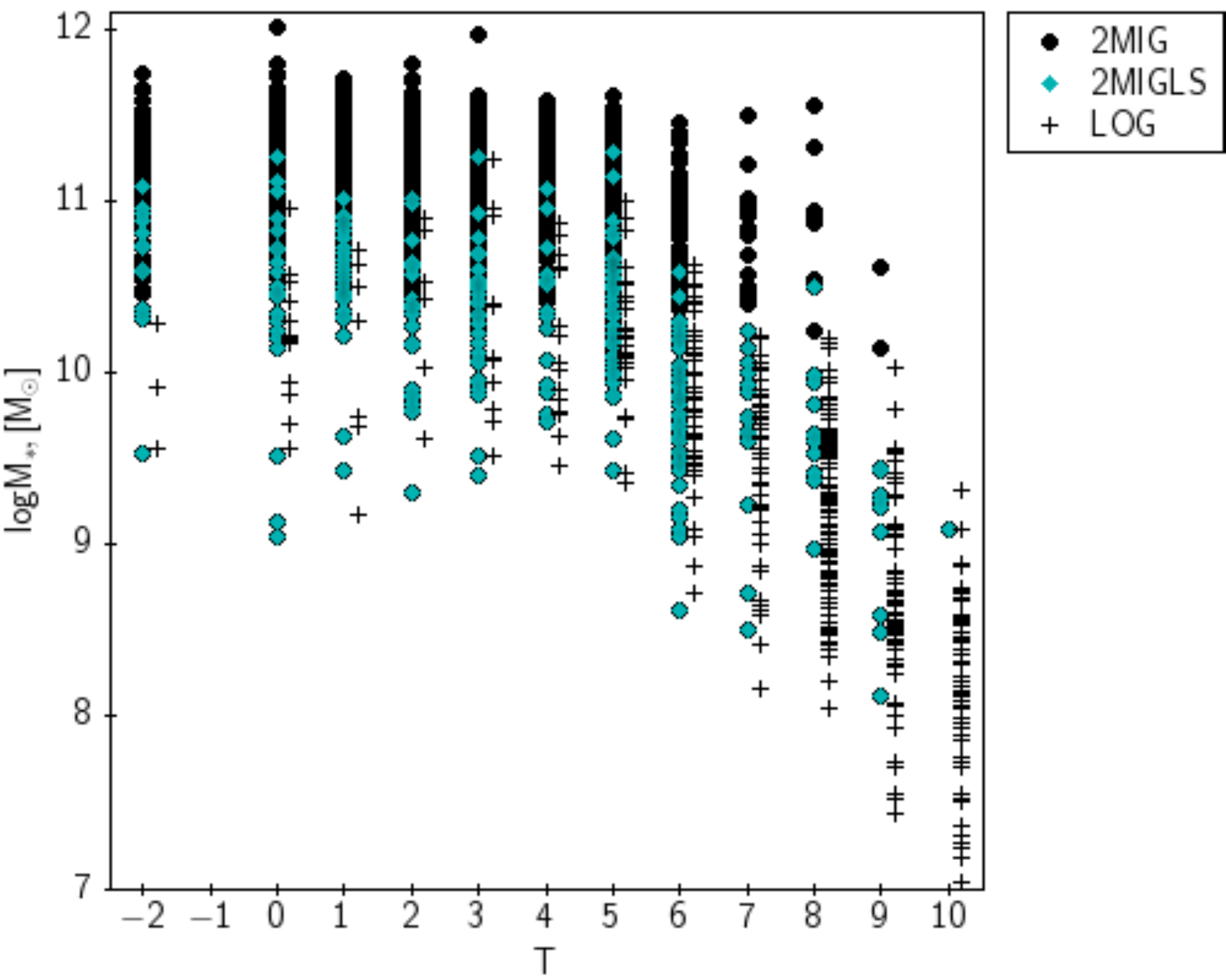} \\
\end{tabular}
\caption{Left: the morphological type distribution for the full 2MIG (N=1616), 2MIG LS ($V_{LG}<$3500 km s$^{-1}$, N=258) and LOG (N=428) samples; 
Middle: stellar mass distribution; Right: dependence of stellar mass on morphological galaxy type.}
\label{2}
\end{figure*}

\begin{figure*}
\tabcolsep 0 pt
\begin{tabular}{cc}
\includegraphics[width=0.5\textwidth,natwidth=350,natheight=250]{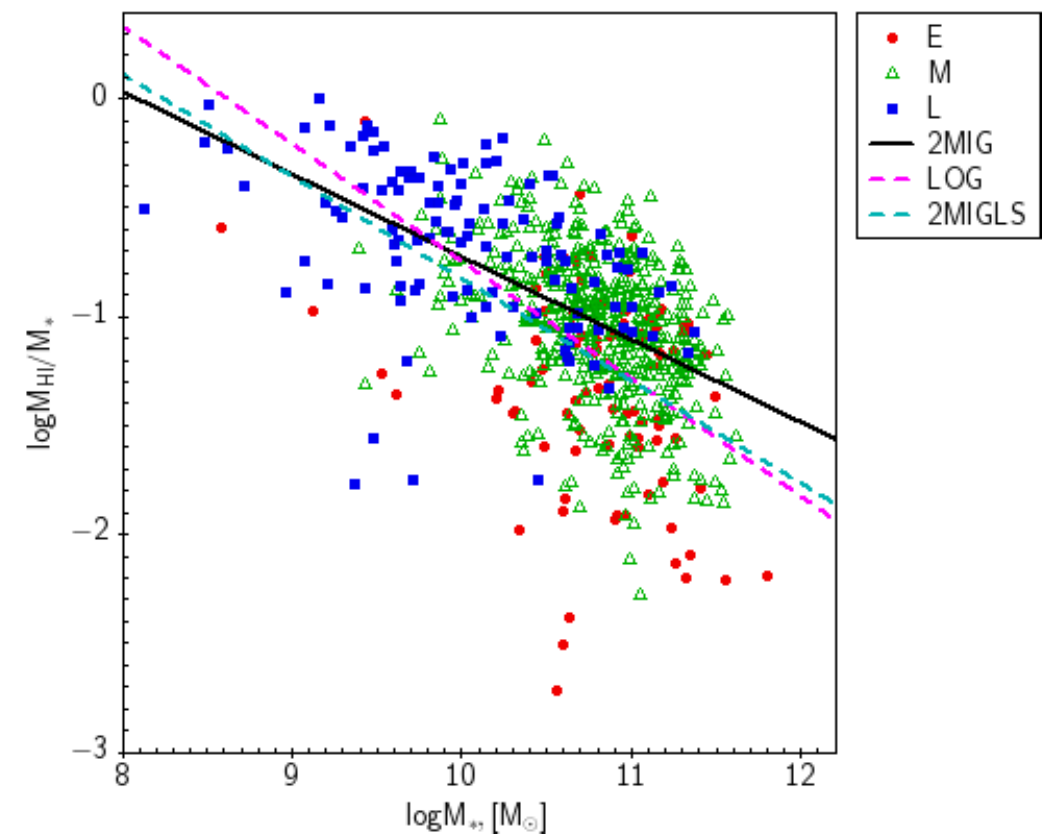} &
\includegraphics[width=0.5\textwidth,natwidth=350,natheight=250]{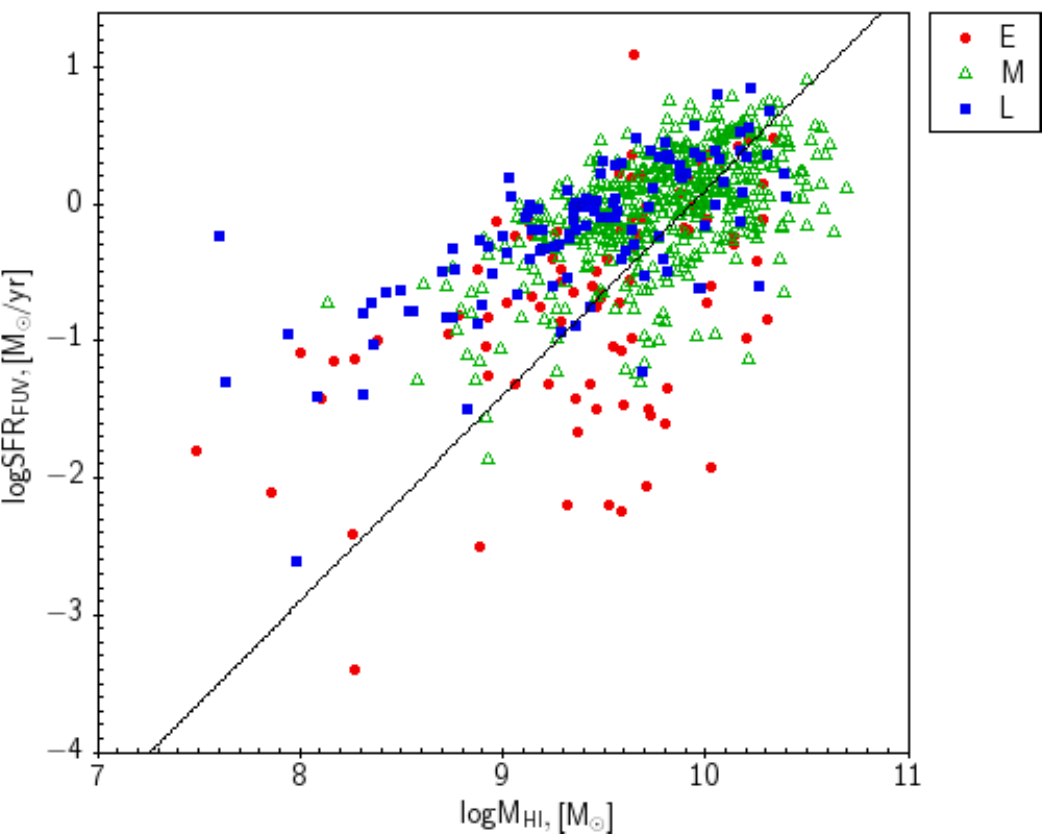} \\
\end{tabular}
\caption{Left: the fraction of neutral hydrogen as a function of stellar mass. The lines show linear regressions for different samples as
noted in figure: log$M_{HI}/M_{*}$=-0.36$\times$log${M_*}$+2.87, log$M_{HI}/M_{*}$=-0.54$\times$log${M_*}$+4.64  and 
log$M_{HI}/M_{*}$=-0.47$\times$log${M_*}$+3.87 for 2MIG (N=663), LOG (N=473) and 2MIG LS (N=268), respectively. Right: the star formation 
rate as a function of hydrogen mass, the line has a slope of 3/2, representing the Schmidt-Kennicutt law (Kennicutt 1998). The morphological 
type division corresponds to E (-2$\leq$T$\leq$1), M (2$\leq$T$\leq$5) and L (6$\leq$T$\leq$10) types.}
\label{3}
\end{figure*}

\subsection {The photometry and formulas}

The current 2MIG and LOG samples consist of 1616 and 428 galaxies, respectively. These samples were observed with the GALEX 
telescope\footnote{Galaxy Evolution Explorer: http://www.galex.caltech.edu/} in far ultraviolet 
($FUV$, $\lambda_{eff}$ = 1539 \AA, FWHM = 269 \AA) wavelength. The photometry was taken 
from the GALEX pipeline All-Sky catalogue\footnote{http://galex.stsci.edu/GR6/; \\ http://galex.stsci.edu/GalexView/}, 
while the photometry for the nearest large diameter galaxies were taken from Gil de Paz et al. (2007), 
Karachentsev, Makarov \& Kaisina (2013) and Karachentsev et al. (2013). 
Since UV wavelengths are direct tracers of young, massive stars and FUV flux is one the most reliable indicators of the $SFR$ 
(Lee et al. 2009), the global rate of star formation in a galaxy, $SFR_{FUV}$, was calculated according to Kennicutt et al. (1998; 
see also Salim et al. 2007, Lee at al. 2009):

\begin{equation}
SFR_{FUV}(M_{\odot} yr^{-1}) = 1.4 \cdot 10^{28} L_{\nu} (FUV), 
\end{equation}
where $L_{\nu}$ is expressed in erg s$^{-1}$ Hz$^{-1}$. 

It is known that mid-infrared (IR) emission is an indirect tracer of $SFR$. This is due to re-emmission starlight on small-grain dust 
(for details see review by Calzetti 2012). We took into account the IR impact into $SFR$ using the WISE\footnote{Wide field Infrared Survey 
Explorer: \\ http://irsa.ipac.caltech.edu/applications/wise/}  
W4 (22 $\mu$m) mid-IR band from the pipeline 
catalogue. At first we calculated $SFR_{IR}$ calibrated by Jarrett et al. (2013):

\begin{equation}
SFR_{IR}(M_{\odot} yr^{-1})=7.50 \cdot 10^{-10} {\nu}L_{22}(L_{\odot}). 
\end{equation}

We then defined the total (FUV+IR) star formation rate, $SFR_{tot}$ as in Jarrett et al. (2013):

\begin{equation}
SFR_{tot}=0.83 \cdot SFR_{IR}+SFR_{FUV}.
\end{equation}

Since the stellar population of galaxies has an average mass-to-luminosity ratio in the $K$-band M/L$\sim$1 
(Bell et al. 2003, see also Jarrett et al. 2013), the stellar masses of galaxies were derived from $K_s$ luminosity. A similar approach was 
also applied in previous works by Karachentsev, Makarov \& Kaisina (2013) Karachentsev \& Kaisina (2013), Karachentsev et al. (2013) 
and Karachentseva et al. (2014) with which we compare our results\footnote{In the listed papers the $K$-band galaxy magnitudes were 
derived from $B$-band corresponding magnitudes. This was due to underestimation of $Ks$ magnitudes for very local big size objects in the 2MASX catalogue. However,
the $K_s$ total magnitude estimation gives a slightly larger masses than using $B$-band. We therefore keep in mind that calculation of the masses 
 using $B$-band calibration includes larger masses for the 2MIG galaxies in 0.1-0.2dex. In other words, from Fig. 2, it
follows that if the stellar masses for the LOG galaxies were calculated from $Ks$ total, their masses would be shifted further to the
low end of the mass function.}.
In this work we used the total $Ks$-band magnitudes from 2MASX catalogue, the $Ks$ absolute magnitude of Sun is 3.32 (Jarrett et al. 2013).

The specific star formation rate $SSFR$ is a $SFR$ per mass unit, expressed in $yr^{-1}$: 

\begin{equation}
SSFR=SFR/M_*.
\end{equation}

\begin{figure*}
\tabcolsep 0 pt
\begin{tabular}{ccc}
\includegraphics[width=0.35\textwidth,natwidth=500,natheight=380]{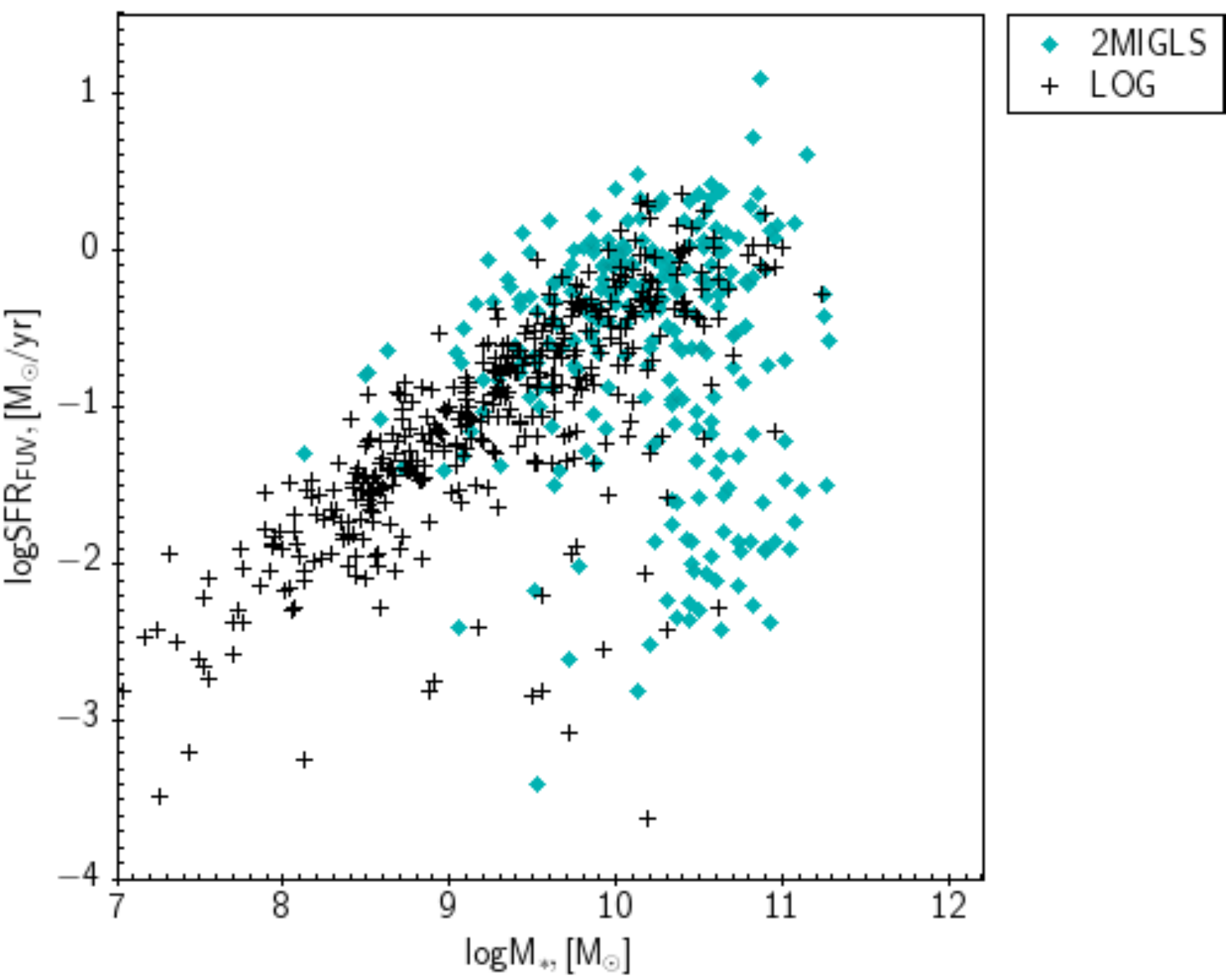} &
\includegraphics[width=0.35\textwidth,natwidth=500,natheight=380]{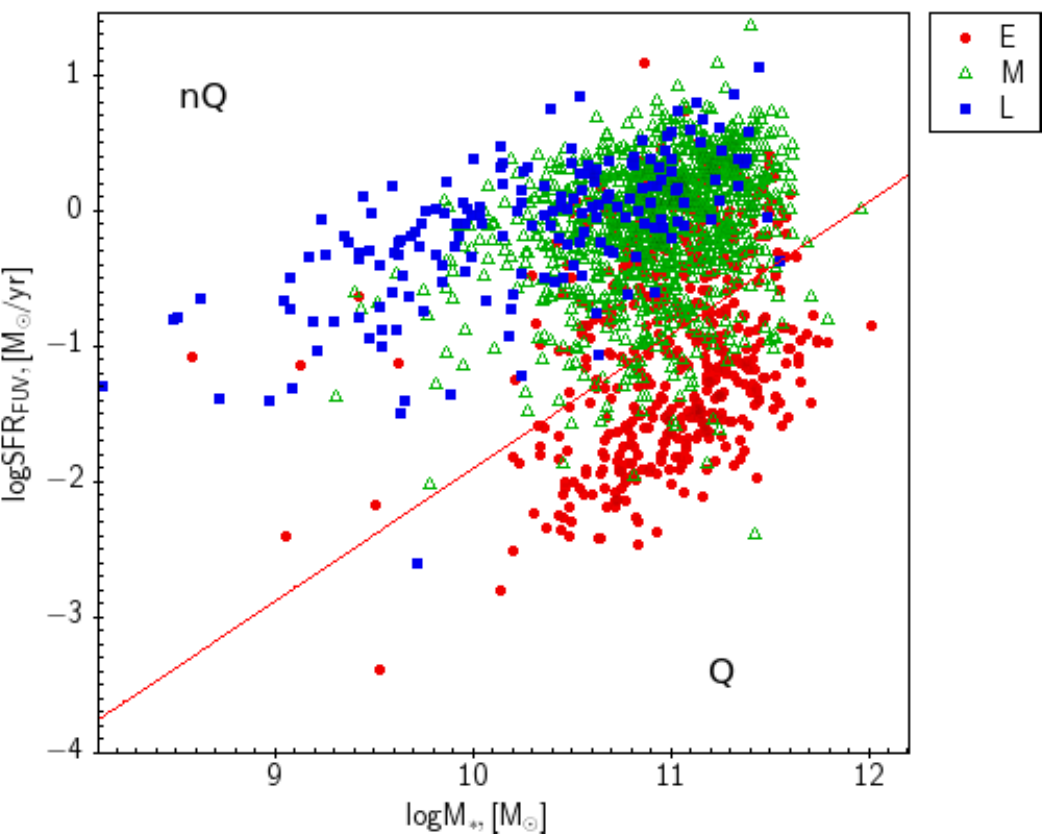} &
\includegraphics[width=0.35\textwidth,natwidth=500,natheight=380]{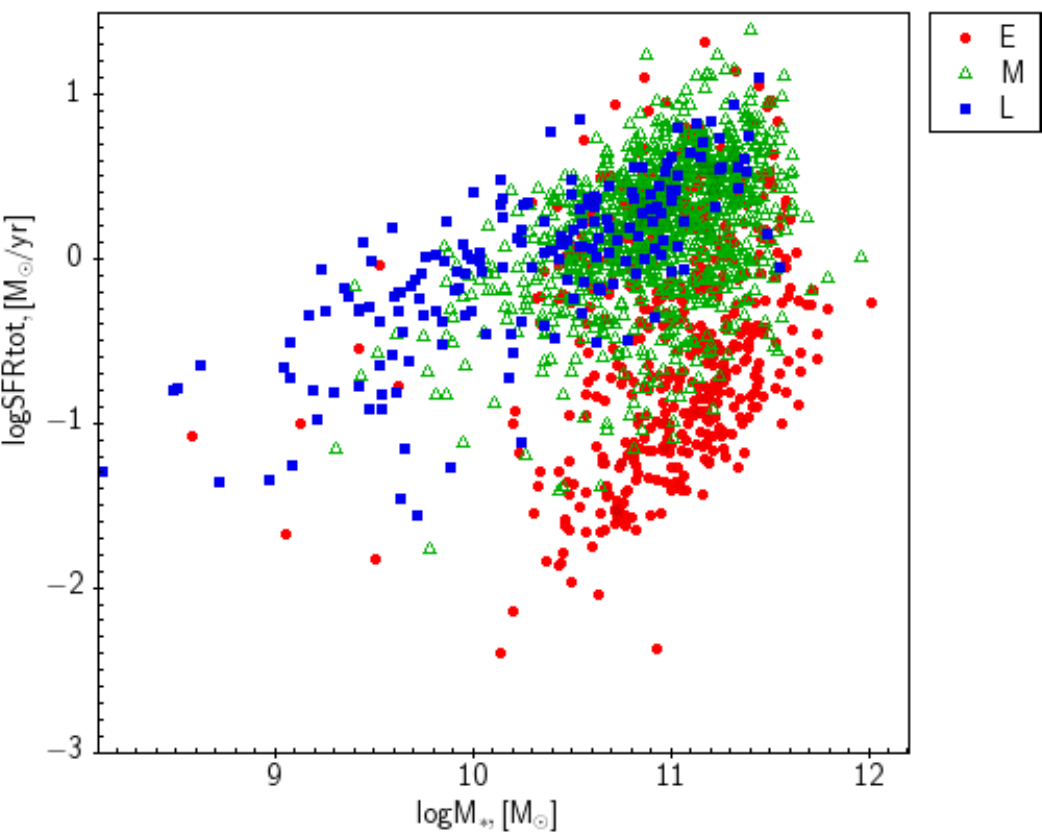} \\
\end{tabular}
\caption{Left: the star formation rate (FUV) as a function of stellar mass for the 2MIG LS and LOG samples, N=258 vs. N=428, respectively.
Middle: the star formation rate (FUV) as a function of stellar mass for the full 2MIG sample, N=1616. 
Right: the total (FUV+IR) star formation rate vs. stellar mass for the full 2MIG sample. The morphological type division corresponds
to E (-2$\leq$T$\leq$1), M (2$\leq$T$\leq$5) and L (6$\leq$T$\leq$10) types. The red line divides the quenched (Q) and non-quenched (nQ) 
sub-samples corresponding to the $FUV-K$ (AB)=6.6 threshold.}
\label{4}
\end{figure*}

Galactic extinction for $Ks$ and $FUV$ bands was calculated using $E(B - V)$ values according to the Schlegel et al. (1998) maps and 
the Cardelli et al. (1989) extinction law where $Rv$=3.1. Internal absorption in galaxies was taken into account with values of 0.085$\times a_{i}$ and 
1.93$\times a_{i}$ for $Ks$ and $FUV$ bands respectively. The coefficients were taken from Karachentsev, Makarov \& Kaisina (2013) and the value of $a_{i}$ -- 
the internal galaxy extinction in $B$-band taken from the Hyperleda. The WISE W4 band was neither corrected for neither Galactic nor internal extinctions.

The hydrogen mass of the galaxy was calculated using the formula from Roberts \& Haynes (1994):

\begin{equation}
M_{HI} = 2.356 \cdot 10^5 \cdot D^2 \cdot F_{HI},
\end{equation}

where $M_{HI}$ is expressed in solar masses, $D=V_{LG}/H_{0}$ in Mpc, $V_{LG}$ is a galaxy radial velocity corrected for the motion of the 
Local Group centroid with apex parameters taken from Karachentsev \& Makarov (1996) and $H_{0}$=72 km s$^{-1} Mpc^{-1}$. The flux $F_{HI}$
in Jy km $s^{-1}$ was calculated as log$F_{HI}$=0.4$^{(17.4-m_{21c})}$, where $m_{21c}$ is a magnitude corrected for galaxy inclination 
taken from the Hyperleda. The $m_{21c}$ data was available only for 663 2MIG galaxies (41\%) and 389 LOGs (91\%) of the samples with FUV data.

The evolutionary state of the galaxy was characterized by the dimensionless parameters $P$ (Past) and $F$ (Future) proposed by
Karachentsev \& Kaisin (2007):

\begin{equation}
P=log(SFR \times T_0/L_K),
\end{equation}

\begin{equation}
F=log(1.85 \times M_{HI}/SFR \times T_{0}).
\end{equation}

It should be noted that the parameter $P$  represents $SSFR$ taken over the entire age 
of the Universe, $T_0$ = 13.7 Gyr. The $F$ parameter corresponds to
the notion of gas depletion time, expressed in units of $T_0$. The
coefficient 1.85 at $M_{HI}$ takes into account the
contribution of helium and molecular hydrogen to the total mass
of the gas (Fukugita \& Peebles 2004).

Throughout the paper, the 2MASX $Ks$ and the WISE W1 (3.4 $\mu$m), W2 (4.6 $\mu$m), W3 (12 $\mu$m) and W4 (22 $\mu$m) magnitudes 
are in the Vega system, except cases where the AB magnitude system is specified. The GALEX $FUV$ magnitudes are always in the AB system.

\section{Star formation properties of isolated galaxies from the 2MIG and LOG samples}

\subsection{Types and masses}

\begin{table*}
\caption{The mean with standard deviation ($SD$) and median values in the quartile range ($Q$) of star 
formation and specific star formation rates in the 2MIG galaxies. The morphological type division 
corresponds to E (-2$\leq$T$\leq$1), M (2$\leq$T$\leq$5) and L 
(6$\leq$T$\leq$10) types, Q for the quenched (FUV-K (AB)$>$6.6) and nQ for the non-quenched (FUV-K (AB)$<$6.6) galaxies.} 
\tabcolsep 4 pt
\begin{tabular}{lccccccccc} \hline
Sample  & \multicolumn{3}{c}{log$SFR_{FUV}$, [M$_{\odot}$/yr]} & \multicolumn{2}{c}{log$SSFR_{FUV}$, [M$_{\odot}$/yr]} & 
\multicolumn{2}{c}{log$SFR_{tot}$, [yr$^{-1}$]} & \multicolumn{2}{c}{log$SSFR_{tot}$, [yr$^{-1}$]}\\
 \hline
 & N & Mean$\pm$SD & Median$\pm$Q & Mean$\pm$SD & Median$\pm$Q  & Mean$\pm$SD & Median$\pm$Q & Mean$\pm$SD & Median$\pm$Q \\
  \hline
  All & 1616 &  -0.38$\pm$0.71 & 0.21$^{+0.37}_{-0.61}$ & -11.26$\pm$0.80 & -11.13$^{+0.41}_{-0.65}$ &  -0.04$\pm$0.61 & 0.08$^{+0.32}_{-0.48}$ & -10.92$\pm$0.65 & -10.78$^{+0.30}_{-0.54}$ \\
 E & 470 &  -1.04$\pm$0.71 & -1.08$^{+0.60}_{-0.49}$ & -12.05$\pm$0.70 & -12.15$^{+0.67}_{-0.48}$ &  -0.51$\pm$0.68 & -0.52$^{+0.53}_{-0.48}$ & -11.51$\pm$0.68 & -11.65$^{+0.67}_{-0.42}$ \\
 M & 969 &  -0.10$\pm$0.50 & -0.01$^{+0.27}_{-0.35}$ & -11.04$\pm$0.53 & -10.97$^{+0.27}_{-0.37}$ &  0.19$\pm$0.43 & 0.23$^{+0.25}_{-0.30}$ & -10.75$\pm$0.44 & -10.69$^{+0.21}_{-0.31}$ \\
 L & 177 &  -0.14$\pm$0.53 & -0.06$^{+0.27}_{-0.34}$ & -10.42$\pm$0.55 & -10.40$^{+0.39}_{-0.40}$ &  -0.04$\pm$0.51 & 0.02$^{+0.29}_{-0.35}$ & -10.32$\pm$0.47 & -10.34$^{+0.34}_{-0.29}$ \\
 \hline
Q & 341 &  -1.42$\pm$0.47 & -1.41$^{+0.34}_{-0.33}$ & -12.49$\pm$0.33 & -12.50$^{+0.29}_{-0.24}$ &  -0.79$\pm$0.55 & -0.82$^{+0.38}_{-0.35}$ & -11.86$\pm$0.49 & -11.95$^{+0.26}_{-0.24}$ \\
nQ & 1275 &  -0.10$\pm$0.46 & -0.05$^{+0.29}_{-0.32}$ & -10.94$\pm$0.52 & -10.93$^{+0.32}_{-0.36}$ &  0.16$\pm$0.44 & 0.21$^{+0.31}_{-0.26}$ & -10.67$\pm$0.43 & -10.65$^{+0.25}_{-0.29}$ \\
 \hline
\end{tabular}
\end{table*}

In this section we describe the basic sample distributions of the 2MIG galaxies 
in comparison with the LOG isolated galaxy catalogue. These were selected 
from totally different galaxy samples applying different selection algorithms (see subsections 2.1 and 2.2.).
Fig. 1 presents the radial velocity distribution of the 2MIG sample. Note that the LOG sample is limited by the Local Supercluster (LS) 
volume $V_{LG}<$3500 km/s. Fig. 2 compares the 2MIG's and LOG's morphological type and stellar mass distributions (on the left and middle panels, respectively) and 
dependence of stellar mass on galaxy morphological type (on the right). We also show in Fig. 2 the same dependences for the 2MIG sample in the same volume as the 
LOG samples (2MIG LS). As can be seen, the galaxies in the LOG catalogue have significantly latter morphological types than the 2MIG galaxies. 
They are up to two-orders of magnitude less massive than the 2MIG galaxies even within the same morphological type. 

We see that different selection algorithms applied to 2D and 3D primary samples define different populations of isolated galaxies. The population of the 
LOG galaxies mostly consists of low mass spiral and late type galaxies which are preferably located in globally low density regions, 
in particular in voids (Elyiv et al. 2013). The population of the 2MIGs, in contrast, is 
totally different: it consists of normal mass, companion-less galaxies of predomintantly early type spirals. These are located in local regions of low density.
From Fig. 2, follows that galaxies in the volume limited 2MIG LS sample are significantly massive that the LOG galaxies. Moreover, we have only 
16 coincident 2MIG LS-LOG galaxies (6\% from the 2MIG LS sample).

\subsection{Star formation rates and neutral hydrogen fraction}

\begin{figure}
\includegraphics[width=0.45\textwidth,natwidth=500,natheight=380]{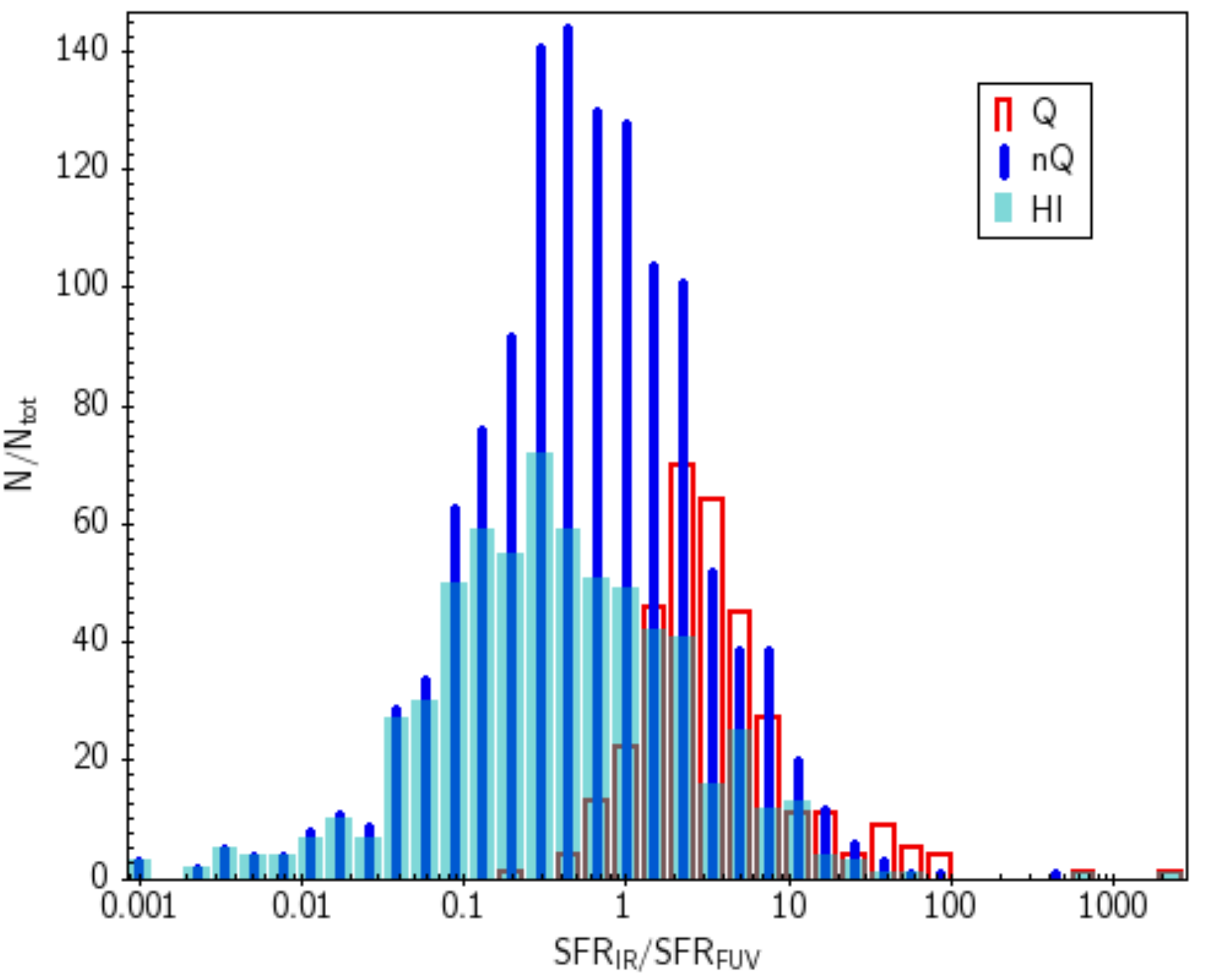} 
\caption{Distribution of $SFR_{IR}$ and $SFR_{FUV}$ ratios for quenched (Q) and non-quenched (nQ) 2MIG galaxies, 
N=341 and N=1275, respectively. 
HI detected galaxies of all morphological types are also shown (N=663).}
\label{5}
\end{figure}

\begin{figure*}
\tabcolsep 0 pt
\begin{tabular}{ccc}
\includegraphics[width=0.35\textwidth,natwidth=500,natheight=380]{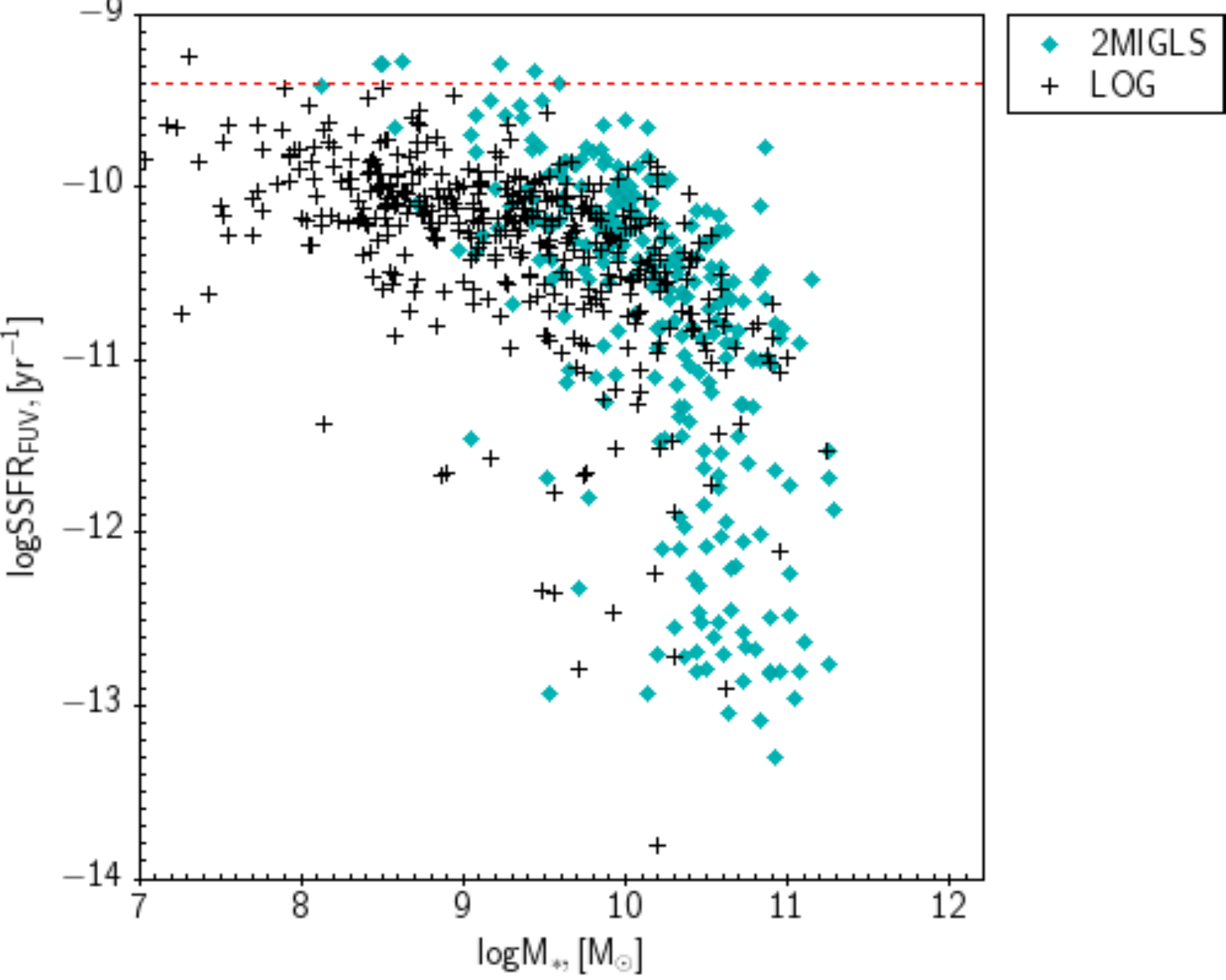} &
\includegraphics[width=0.35\textwidth,natwidth=500,natheight=380]{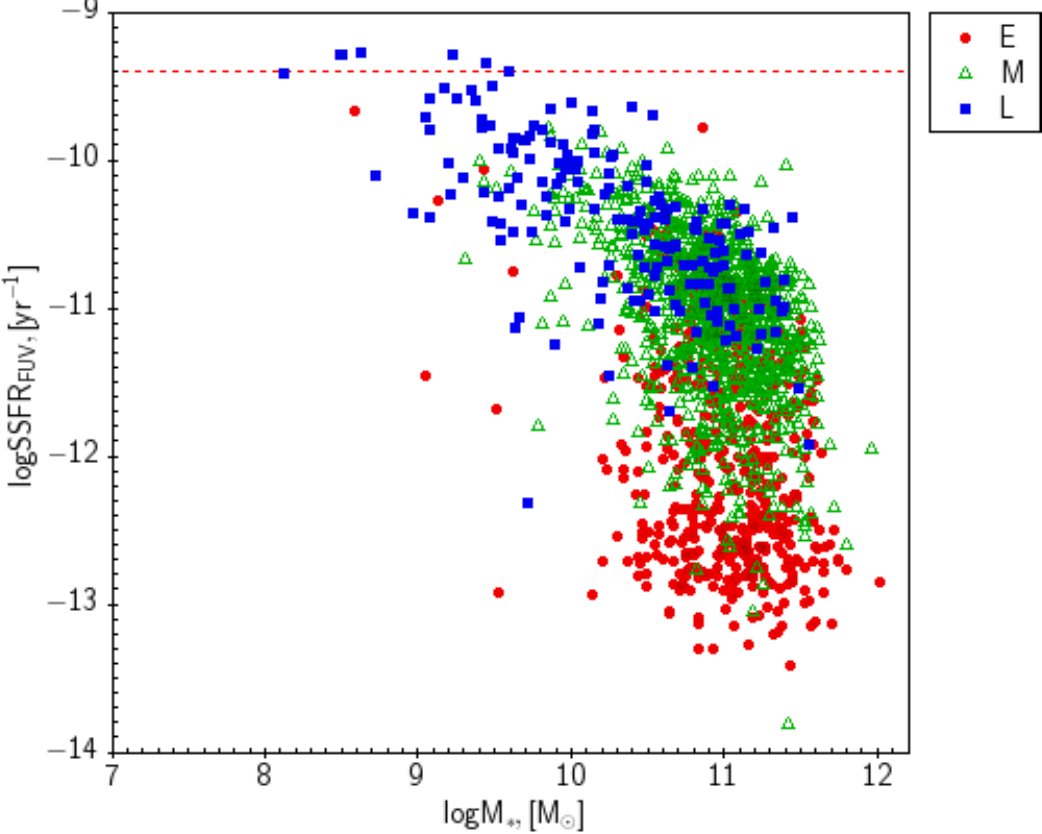} &
\includegraphics[width=0.35\textwidth,natwidth=500,natheight=380]{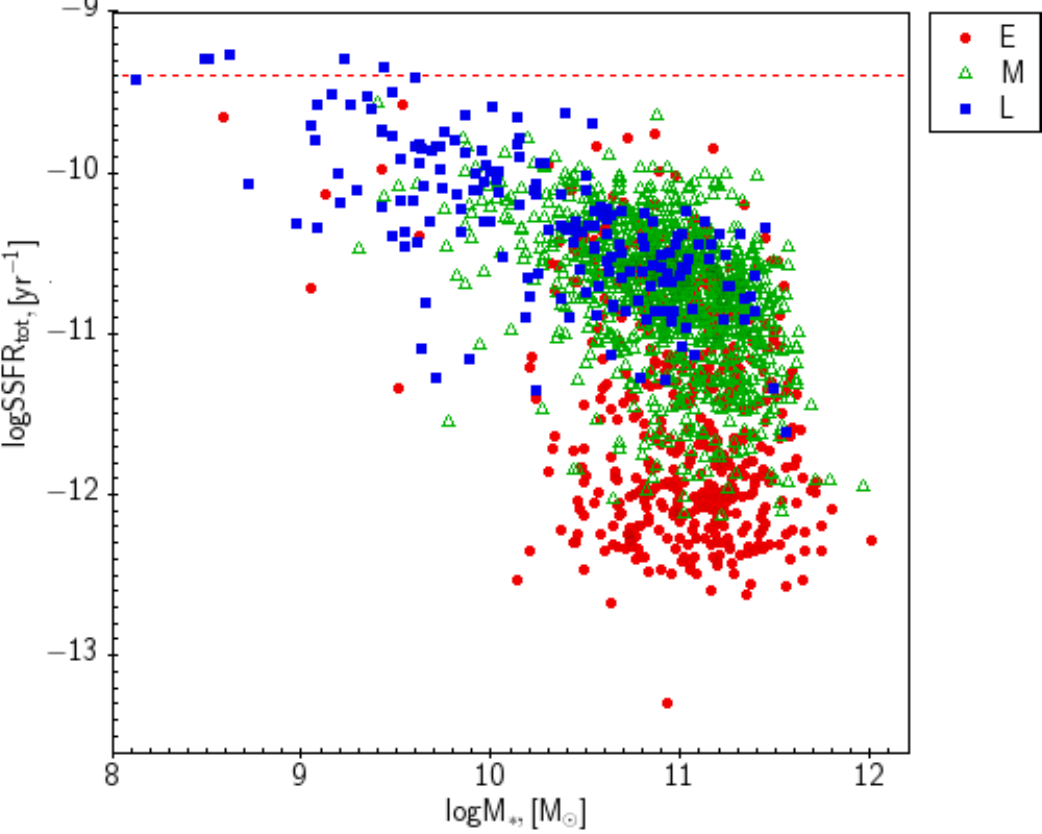} \\
\end{tabular}
\caption{Left: the specific star formation rate (FUV) as a function of stellar mass for the 2MIG LS and LOG samples, N=258 vs. N=428, 
respectively.
Middle: the specific star formation rate (FUV) as a function of stellar mass for the full 2MIG sample, N=1616. 
Right: the total (FUV+IR) specific star formation rate vs. stellar mass for the full 2MIG sample. The morphological type division corresponds
to E (-2$\leq$T$\leq$1), M (2$\leq$T$\leq$5) and L (6$\leq$T$\leq$10) types. The red dashed line corresponds to 
log$SSFR=$-9.4 threshold.}
\label{6}
\end{figure*}

\begin{figure*}
\tabcolsep 10 pt
\begin{tabular}{cc}
\includegraphics[width=0.35\textwidth,natwidth=450,natheight=400]{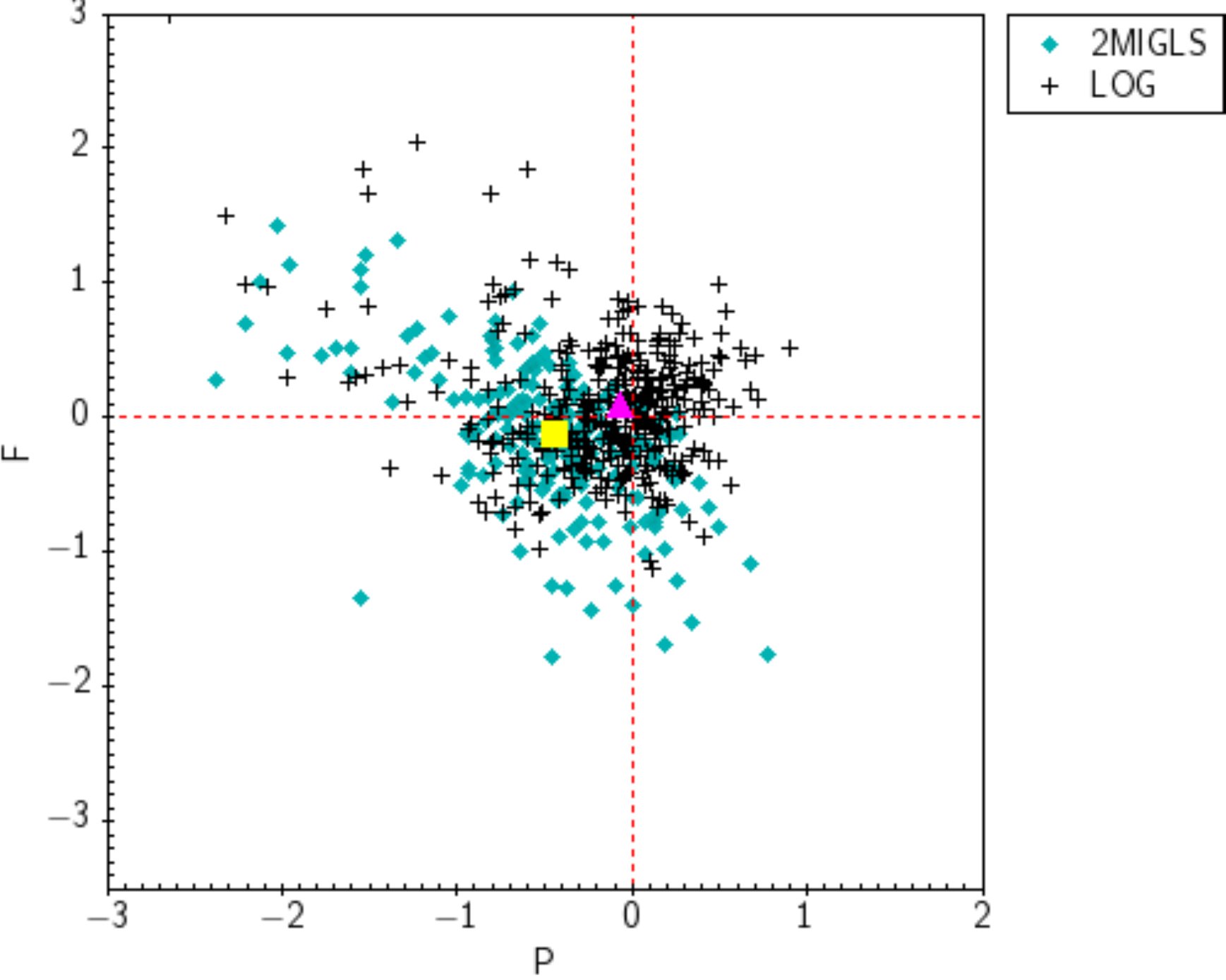} &
\includegraphics[width=0.35\textwidth,natwidth=450,natheight=400]{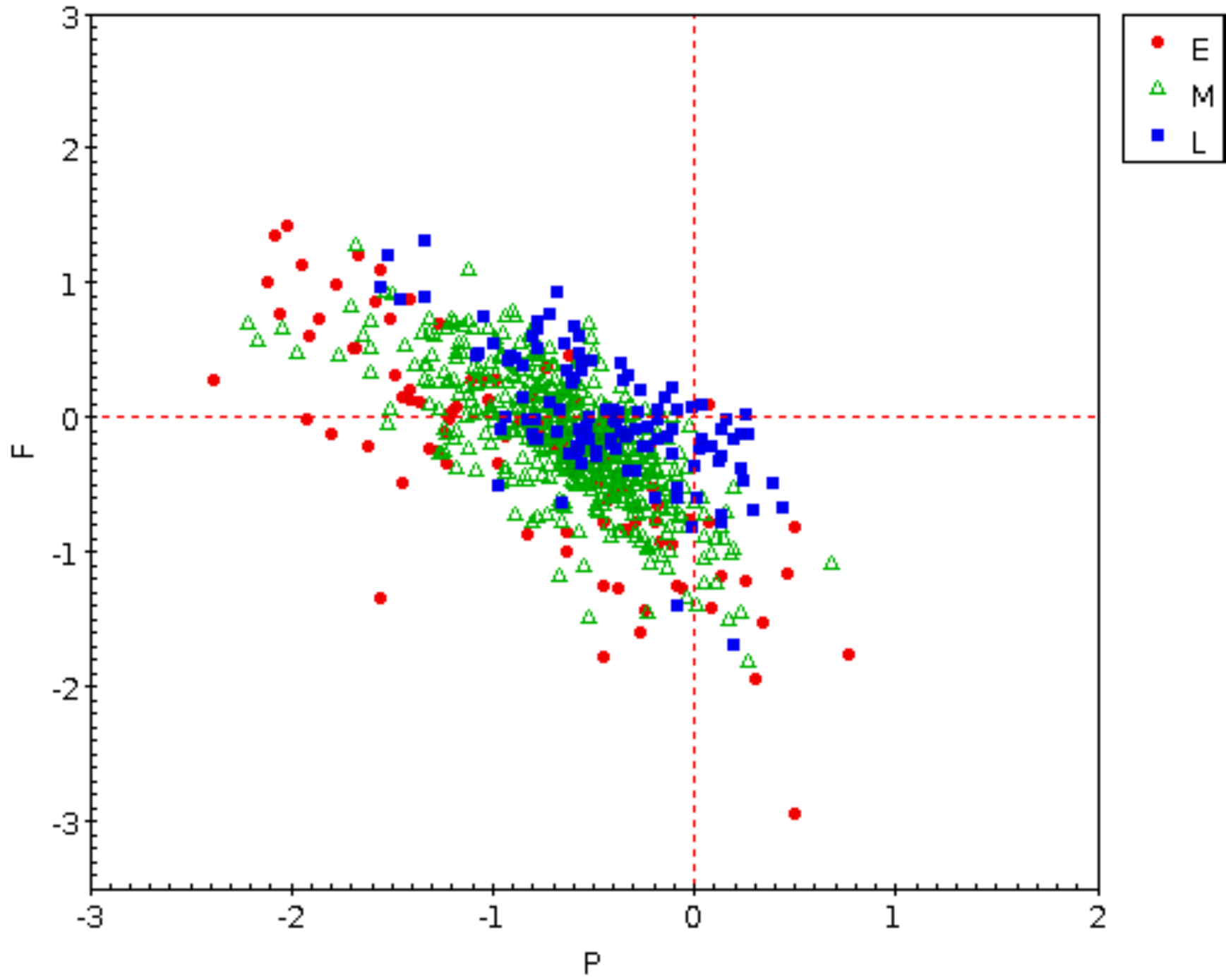} \\
\end{tabular}
\caption{Left: Future vs. Past diagnostic diagram for the 2MIG LS and LOG samples (N=201 and 389, respectively). 
Corresponding median values for the 2MIG LS and LOG samples are marked with square and triangle, respectively. The galaxy with coordinates $P$=0 and $F$=0 is able to 
reproduce its stellar mass during the Hubble time at the currently observed star formation rate. Here, the reserves of gas are
sufficient to support the observed $SFR$ during the alternative Hubble time. Therefore, in general, the LOG galaxies have larger gas reserves 
than 2MIGs. Right: Future vs. Past diagnostic diagrams
for the full 2MIG sample (N=663). The morphological type division corresponds to E (-2$\leq$T$\leq$1), M (2$\leq$T$\leq$5) and L 
(6$\leq$T$\leq$10) types. The P and F mean and median values for the 2MIG galaxies of different types are noted in Table 3. }
\label{7}
\end{figure*}
  
Fig. 3 (left panel) shows  the fraction of neutral hydrogen as a function of stellar mass. 
The fraction of neutral hydrogen increases with decreasing galaxy mass. The lines represent the linear regressions for three samples
2MIG, LOG and 2MIG LS. We see that all slopes are similar, however the relation for the 2MIG sample is flatter because of a smaller 
fraction of galaxies with high fraction of neutral hydrogen and high fraction of galaxies with moderate and higher stellar masses. The slope
for the 2MIG LS sample is a somewhat steeper. The E, M and L morphological type division has the effect of separating
the total morphological scale into 3 groups: early type galaxies (E) -2 -- 1, spiral ``middle'' types (M) 2--5 and late types (L) 6--10. In total, 20\%, 47\% and 66\% 
of our E, M and L galaxies, respectively, are detected in HI meaning that these galaxies are still have significant gas reservoirs. Note, that 
only 663 2MIG galaxies and 389 LOG galaxies in the Hyperleda have HI measurements, 41\% and 91\% from the total 
numbers, respectively.

On the right panel of Fig. 3 we plotted the star formation rate (FUV) as a function of hydrogen mass for the 2MIG galaxies. The line 
represents the well-known Schmidt-Kennicutt law with slope $\alpha$=3/2 (Kennicutt 1998). We see that 
galaxies of all types are located very symmetrically, however the scatter for early type galaxies are larger. 
The existence of early type galaxies with high $SFR_{FUV}$ and small hydrogen mass may tell us that their UV light may originate not only 
from the young stars.

Fig. 4 presents the star formation rate $SFR_{FUV}$ as a function of stellar mass for the 2MIG LS and LOG samples (left panel) and
full 2MIG sample with E, M and L united morphological type division (middle panel). The right panel the same, excepting that total star formation rate $SFR_{tot}$ calculated 
by formula (7). This was done using a linear  combination of $SFR_{FUV}$ and $SFR_{IR}$. From the comparison of the LOG and 2MIG LS 
samples we conclude that the LOG population ``completes'' log$SFR_{FUV}$ vs. logM$_{*}$ main sequence to the low mass end. In each figure we
clearly see the separation between two concentrations: the main sequence and so called ``quenched'' or ``red and dead'' galaxies. We will refer 
to these concentrations as to non-quenched (nQ) and quenched (Q) galaxies in order to have a non-visual classification. 
The border between these two regions may be roughly defined as $FUV-K$ (AB)=6.6  and is marked by the red line 
in Figs. 4 and 8. The nQ concentration consists of mainly M and L galaxies (85\%), however the Q concentration is formed predominantly by
E galaxies (82\%), according to our visually defined types. In total, 95\% of our M+L galaxies are located on the main sequence and, correspondingly,
only 5\% of them are quenched. Meanwhile, 60\% of E type galaxies are located in the Q region while the rest, 40\%, are main sequence, 
i.e. still show signs of star formation. This is confirmed by the fact that at least 
33\% of E type galaxies in nQ region are detected in HI (49\% of all nQ galaxies have HI detections), while only 10\% of E type galaxies in the
Q region still have neutral hydrogen (11\% of all Q galaxies have HI detections). According to Schawinski et al. (2014) this evidences the fact that star formation of 
elliptical galaxies was quenched rapidly
by rapid removal of the gas, contrary to spiral/late type galaxies. In this case, star formation halted far slower due to gradual gas loses driven 
by secular and/or environmental processes.

\subsection{$SFR_{IR}$ vs. $SFR_{FUV}$} 

Table 1 represents the mean and median values of 
log$SFR_{FUV}$, log$SSFR_{FUV}$, log$SFR_{tot}$ and log$SSFR_{tot}$ for the full 2MIG sample and its varying sub-samples. It follows from Table 1 that the ratio between $SFR_{tot}$ 
calculated from $SFR_{IR}$ and $SFR_{FUV}$ using (7), and $SFR_{FUV}$ is somewhat different 
for E, M, L, Q and nQ types: 3.4, 1.9, 1.3, 4.3 and 1.8 times, respectively. The distributions of ratio $SFR_{IR}$/$SFR_{FUV}$ for 
Q, nQ and HI-detected galaxies are shown in Fig. 5. Apparently, our Q (i.e also E) galaxies have higher ratios of $SFR_{IR}$/$SFR_{FUV}$ 
than the other types.
Normally, the ellipticals do not contain much dust, except for giant elliptical galaxies as a result of mergers. 
However, the mid-IR emission in early type galaxies can contain a large contribution (an order of magnitude stronger) 
from intermediate-age and older stellar populations than from the young stars (see references in Salim et al. 2012, Calzetti et al. 2012 and 
Jarrett et al. 2013). In Table 2 we show some characteristics
of galaxies from the 2MIG sample with the highest ratios of $SFR_{IR}$/$SFR_{FUV}>$50. We see that only a few galaxies have SDSS spectra.
However, 2MIG417 and 2MIG1573 have AGN according to the NED. Moreover, these galaxies have high W1-W2 (WISE) IR colours that allow them
to be considered as having bright IR AGN (see subsection 4.2 for the details). 2MIG1155 is a very local galaxy with an enormously high 
$SFR_{IR}$/$SFR_{FUV}=$2689 ratio. The WISE W4 magnitude (0.482) agrees with Hershel 24 $\mu$m MIPs measurement
(5.2 Jy, Bendo, Galliano \& Madden 2012). The galaxy also contains a small fraction of neutral hydrogen.  We assume that 2MIG1155 could be a 
starburst due to the value of $SSFR_{tot}$=-9.58 yr$^{-1}$ and by taking into account its location on colour-colour plot in Fig.12 (starburst 
region, Jarrett et al. 2011). In general, all galaxies listed in Table 2 are located in or near that region: 
see for example the localization of 2MIG1840 and 2MIG842 in Fig. 12 (second and third galaxy with the highest levels of $SFR_{IR}$/$SFR_{FUV}$).

We also note that 2MIG2024 is probably a starforming galaxy in outgoing 
merging with a moderate $SSFR_{tot}$=-10.19 yr$^{-1}$. This is also confirmed by its SDSS spectrum. 2MIG1142, 2MIG1443, 2MIG1840 and 2MIG1987 
have signs of disturbance and/or visible small satellites. These could have weak influence on main galaxy properties (Karachentseva et al. 2011, Melnyk et al. 2014).  

With current data, we are not able to assert the starforming/starburst origin of high IR emission in all early type galaxies, 
neither may we exclude the AGN origin of IR emission due once again to insufficient data. We have assumed that the $SFR_{tot}$ values are more reliable for the 
spiral galaxies since they normally contain dusty, i.e. starforming, regions, contrary to the early type galaxies.

We therefore present both values of log$SFR_{FUV}$/log$SFR_{tot}$ and 
log$SSFR_{FUV}$/log$SSFR_{tot}$ throughout the paper which show, in general, the same tendencies as in the graphs.

\begin{table*}
\caption{2MIG galaxies with the highest ratio of $SFR_{IR}$/$SFR_{FUV}$.} 
\tabcolsep 2 pt
\begin{tabular}{lcccccccc} \hline
2MIG & Name & T & log$M_{HI}/M_*$ & log$SFR_{tot}$, [$M_{\odot}/yr$] & $SFR_{IR}$/$SFR_{FUV}$ & AGN & AGN$_{W1-W2}$ & Comment \\
  \hline
87 & MCG-02-03-004 &  0 & -- & 0.29 & 99 & -- & 0.07 & n$^3$ \\
417& MCG-02-09-040 &  0 & -- & 0.51 & 96 & Sy2, NED & {\bf 0.62} & n\\
428 & 2MASXJ03304719-6747056 &  0 & -- & -0.21 & 92 & -- & 0.11 & n\\
722 & ESO159-019 &  0 & -- & -0.07 & 54 & -- & 0.17 & n\\
842 & ESO206-013 &  3 & -- & 0.17 & 430 & -- & 0.06 & n\\
1142 & MCG-03-22-002 &  1 & -- & -0.15 & 55 & -- & 0.05 & disturbed? \\
1155 & ESO495-021 &  -2 & -1.26 & -0.05 & 2689 & -- & 0.42 & compact, starburst?\\
1443 & CGCG266-050 &  0 & -- & 0.44 & 96 & n & 0.15 & dusty, disturbed?, 2 sat?$^4$ \\
1573 & 2MASXJ11240275-2823155 &  3 & -- & 0.32 & 74 & AGN, NED & {\bf 1.51} &  n\\
1840 & IC0860 &  0 & -2.72 & 0.72 & 812 & n & 0.14 & 1 small sat + 2 sat?\\
1987 & UGC09307 &  2 & -- & 0.47 & 62 & TO$^{1}$ & 0.23 & 1 small sat\\
2024 & CGCG273-026 &  1 & -- & 1.14 & 58 & SF$^2$ & 0.25 & major merging \\
2412 & UGC10926 &  2 & -- & 0.26 & 50 & -- & 0.13 & 1 small sat projection \\
 \hline
\end{tabular}
$^1$Transition object, located upon the border of a starburst galaxy and AGN, according to Coziol et al. (2011) \\
$^2$Starforming galaxy according to the SDSS. \\
$^3$ ``--'' means no data, ``n'' means no AGN or no special comment on appearance of the object. \\
$^4$ ``sat?'' Small satellite without radial velocity (projected on the sky).
\end{table*}

\subsection{Specific star formation rates}
Fig. 6 presents the specific star formation rate, calculated by formula (8), as a function of stellar mass for the same samples as in Fig. 4.
We see that only 5 galaxies have a little larger log$SSFR$ than -9.4. All of these objects
are very local galaxies with $V_{LG}<$600 km/s: 2MIG405=NGC1311,
2MIG646=NGC1744, 2MIG1699=UGC07321, 2MIG1724=NGC4395\footnote{Pulatova et al. (2015) argued about a presence of the 
significant companions in the neighbourhood of NGC4395 that preclude it from being considered isolated.}. In addition, 2MIG1741=UGC07699 
the $K_s$ catalogue magnitude could be underestimated due to a large diameter ($>$2.5 arcmin). It is interesting that not one of these galaxies is a 
member of the LOG catalogue.
We see that the threshold of log$SSFR$ for 2MIG galaxies is near $\sim$dex(-9.4) [-yr$^{-1}$].
This agrees with results obtained by Karachentsev \& Kaisina (2013) and Karachentsev et al. (2013) for the local normal galaxies.

Fig. 7 shows Future vs. Past diagnostic diagram for the LOG and 2MIG LS samples (left panel) and for the full 2MIG sample (right panel).
In Table 3 we compared the mean with standard deviation ($SD$) and median in the quartile range ($Q$) values of these parameters for 
the full 2MIG sample and its sub-samples.

According to formulas (10-11) P and F parameters are dimensionless. The result is that 
the galaxy with coordinates $P$=0 and $F$=0 is able to reproduce its stellar mass
during the Hubble time at the currently observed star formation rate. It follows that, the reserves of gas are
sufficient to support the observed $SFR$ during a second Hubble time. It follows from the left 
panel of Fig. 7 (see also Karachentsev et al. 2013) that only half of the LOGs'
gas reserves are dissipated. In general, the value of the $P$ parameter for the 2MIG LS galaxies is lower 
than one for the LOGs. As a result, the average current star formation rate is
able to reproduce only a part ($\sim$1/2) of their stellar mass over a time $T_{0}$. Thus,
in past epochs the star formation rate of the 2MIG galaxies was somewhat higher. In particular,  
L-type galaxies have almost the same amount of gas as LOGs during present epochs. However the gas fraction in M and E type galaxies is much smaller 
and will be consumed over less time than $T_0$.

\begin{table}
\caption{Past and Future dimensionless parameters for the 2MIG galaxies. The morphological type division corresponds to E (-2$\leq$T$\leq$1), M (2$\leq$T$\leq$5) and L 
(6$\leq$T$\leq$10) types, Q for the quenched (FUV-K (AB)$>$6.6) and nQ for the non-quenched (FUV-K (AB)$<$6.6) galaxies.} 
\tabcolsep 4 pt
\begin{tabular}{lccccc} \hline
Sample  & \multicolumn{3}{c}{P} & \multicolumn{2}{c}{F} \\
 \hline
 & N & Mean$\pm SD$ & Median$\pm Q$ & Mean$\pm SD$ & Median$\pm Q$ \\
  \hline
  All & 663 &  -0.62$\pm$0.47 & -0.58$^{+0.25}_{-0.29}$ & -0.14$\pm$0.54 & -0.12$^{+0.31}_{-0.30}$ \\
 E & 92 &  -0.83$\pm$0.71 & -0.67$^{+0.38}_{-0.74}$ & -0.27$\pm$0.81 & -0.22$^{+0.48}_{-0.57}$ \\
 M & 455 &  -0.63$\pm$0.40 & -0.60$^{+0.24}_{-0.25}$ & -0.15$\pm$0.47 & -0.13$^{+0.28}_{-0.29}$ \\
 L & 116 &  -0.43$\pm$0.42 & -0.41$^{+0.30}_{-0.27}$ & 0.01$\pm$0.47 & -0.06$^{+0.41}_{-0.16}$ \\
 \hline
 Q & 38 &  -1.41$\pm$0.72 & -1.59$^{+0.55}_{-0.35}$ & 0.09$\pm$0.92 & 0.31$^{+0.38}_{-0.65}$ \\
 nQ & 625 &  -0.57$\pm$0.41 & -0.81$^{+0.25}_{-0.24}$ & -0.15$\pm$0.50 & -0.13$^{+0.27}_{-0.29}$ \\
 \hline
\end{tabular}
\end{table}

\section{Isolated galaxies vs. paired galaxies}

The star forming properties of the 2MIG galaxies were compared with those of compact pairs' members from the same 
lists used in Melnyk et al. (2014). By this comparison we would like to define the environmental influence (if any exists) 
on the $SFR$ properties of galaxies located in denser environments contrasted with isolated ones.
Here we briefly describe the selection procedure applied for the  galaxy pairs searching.

Initially, we extracted from the 2MASS XSC catalogue galaxies with $Ks<$12 and 
$a_K\geq30''$ according to the 2MIG selection criteria, obtaining the same primary volume-limited sample.
Then, we found the closest neighbour for every galaxy 
from the sample, according to the angular separation between galaxies. We took only 2200 of the most tightly spaced pairs and correlated 
the positions of their galaxies with objects from the NED and Hyperleda databases. Finally, we rejected from the consideration: 
1) all pairs without radial velocities for at 
least one pair member, 2) pairs with velocity differences of $dV>$1000 km/s. We then took into account 
only pair members having $FUV$ magnitude. Our final sample contained 1482 paired members. We considered pairs with $dV<$1000 km/s and 
linear projected distance $R<$240 kpc (the full sample; Pairs), the most compact pairs (CP) having $dV<$150 km/s and $R<$50 kpc (median values are $dV=53^{+48}_{-32}$ km/s and 
$R=29^{+10}_{-10}$ kpc) while the wider pairs (WP) with $dV>$150 km/s or $R>$50 kpc (median values are $dV=156^{+151}_{-80}$ km/s and $R=84^{+80}_{-53}$ kpc). 
Here it is noted that we did 
not apply any isolation criteria for selected ``pairs'', our formal criterion suggests that members of ``pairs'' are closer to each other than 
to any other galaxy. We consider paired galaxies located in high density regions in comparison with 2MIG galaxies. However, a 
visual inspection revealed that the closest neighbours to the CP (WP) members are located no closer than in 4(2)$\times R$. Note that we did not take 
into account blue smaller galaxies (with $a_K<30''$) which are located near selected galaxies approximately in 1/6 cases.
In Melnyk et al. (2014) it was shown that the colour properties of WP\footnote{CP sample from this work corresponds to CP3 sample in Melnyk et al. 
(2014) but WP sample corresponds to CP1 sample excluding CP3.} members are close to galaxies located in groups contrary to CP pairs, 
the colours of which are very close to the colours of galaxies in isolated pairs and triplets (Karachentsev 1987, Karachentseva et al. 1979, 
Karachentseva \& Karachentsev 2000) which contain many interacting systems.

\begin{figure*}
\tabcolsep 0 pt
\begin{tabular}{cc}
\includegraphics[width=0.45\textwidth,natwidth=350,natheight=240]{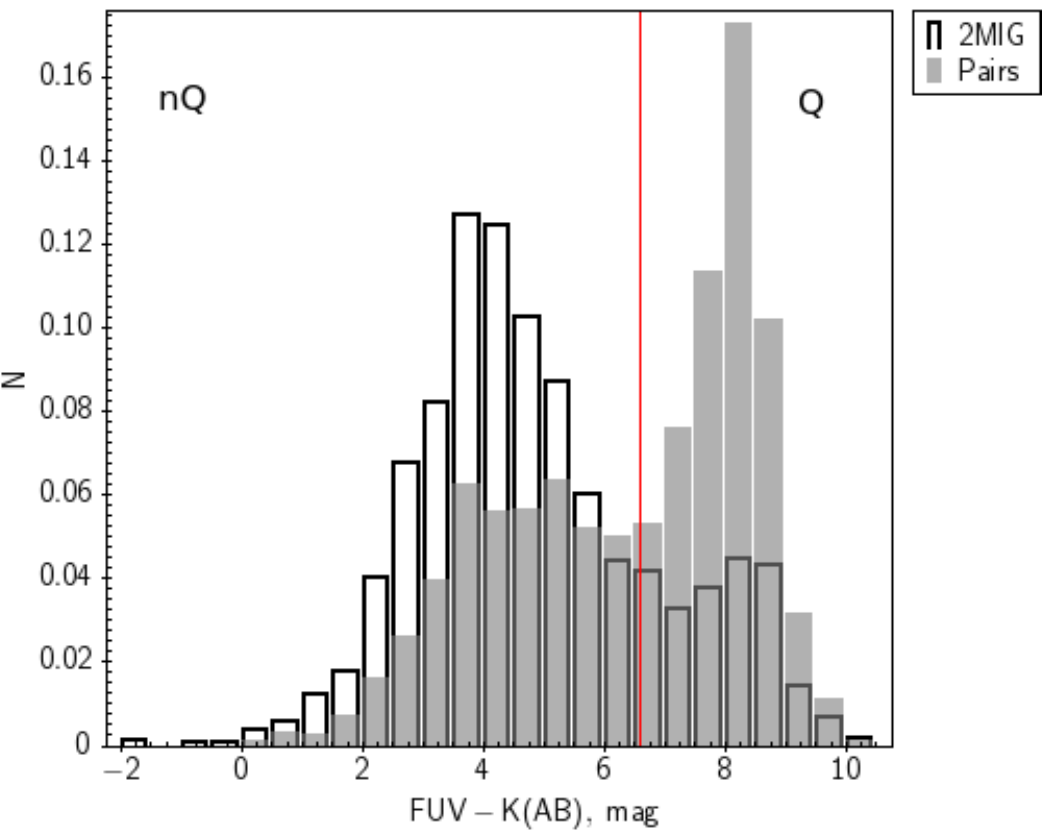} & 
\includegraphics[width=0.45\textwidth,natwidth=350,natheight=240]{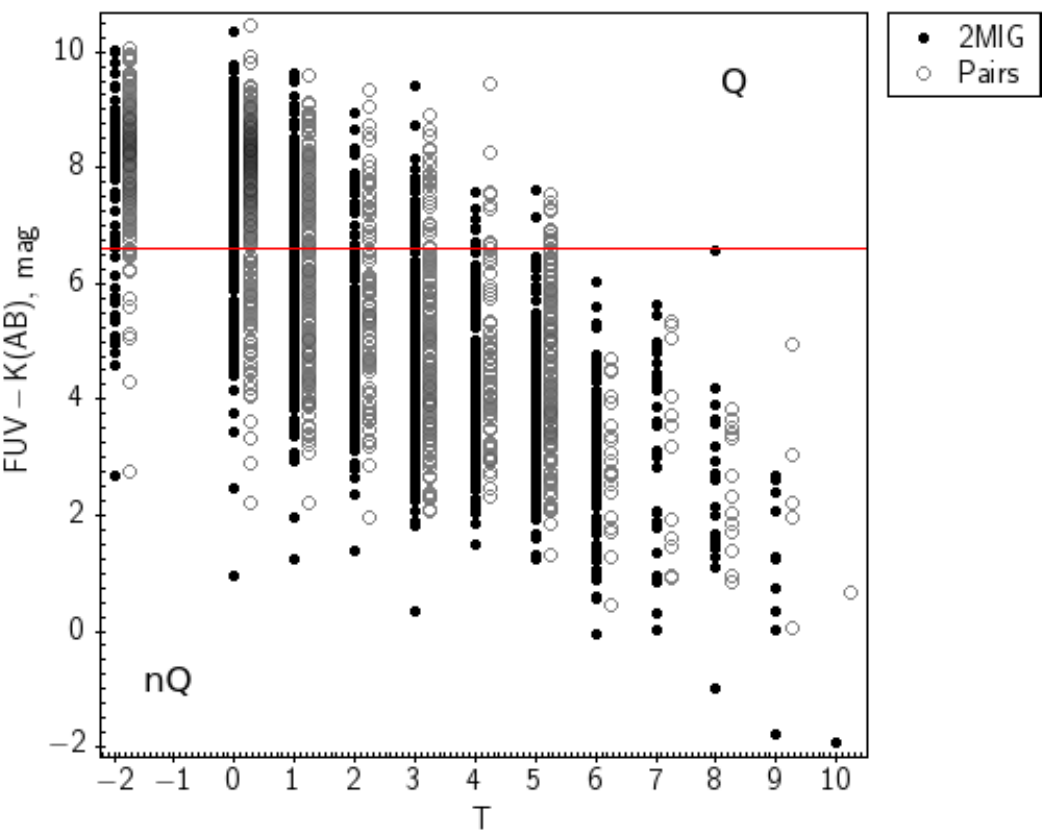} \\
\end{tabular}
\caption{Left: the $FUV-K$ (AB) colour distributions for the 2MIG and paired galaxies. Right: dependence of $FUV-K$ colour on the galaxy 
morphological type.}
\label{8}
\end{figure*}

\begin{figure*}
\tabcolsep 0 pt
\begin{tabular}{cc}
\includegraphics[width=0.45\textwidth,natwidth=350,natheight=240]{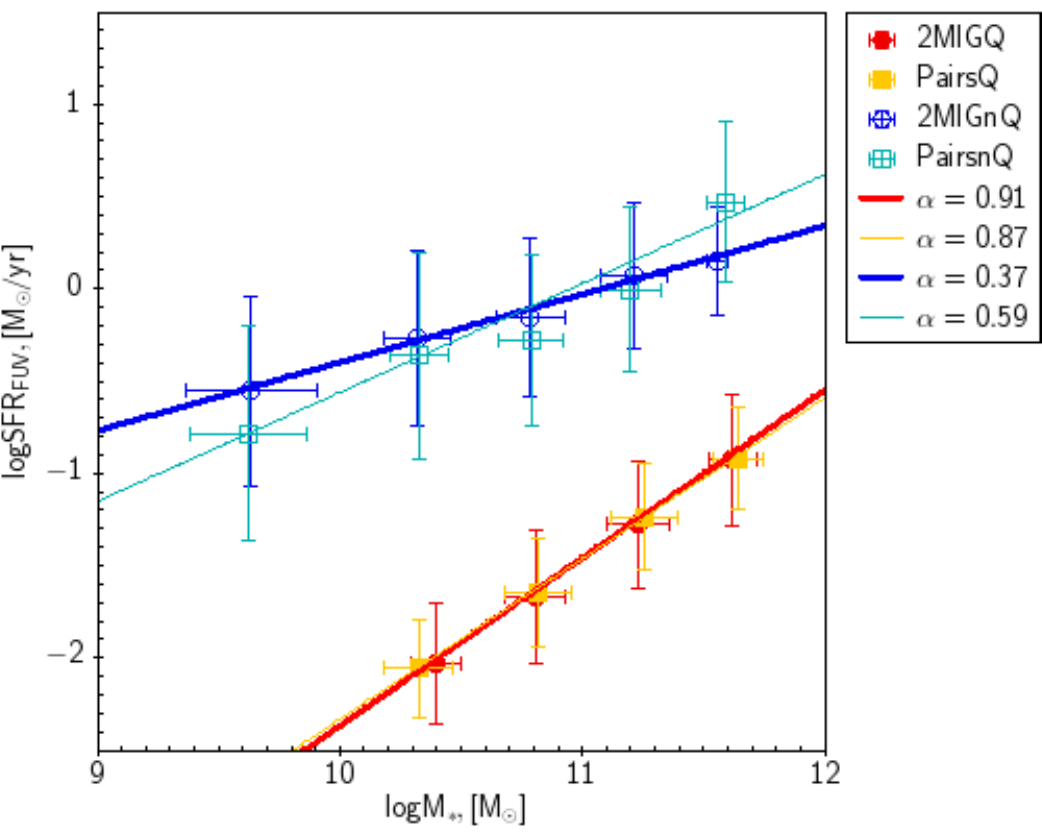} &
\includegraphics[width=0.45\textwidth,natwidth=350,natheight=240]{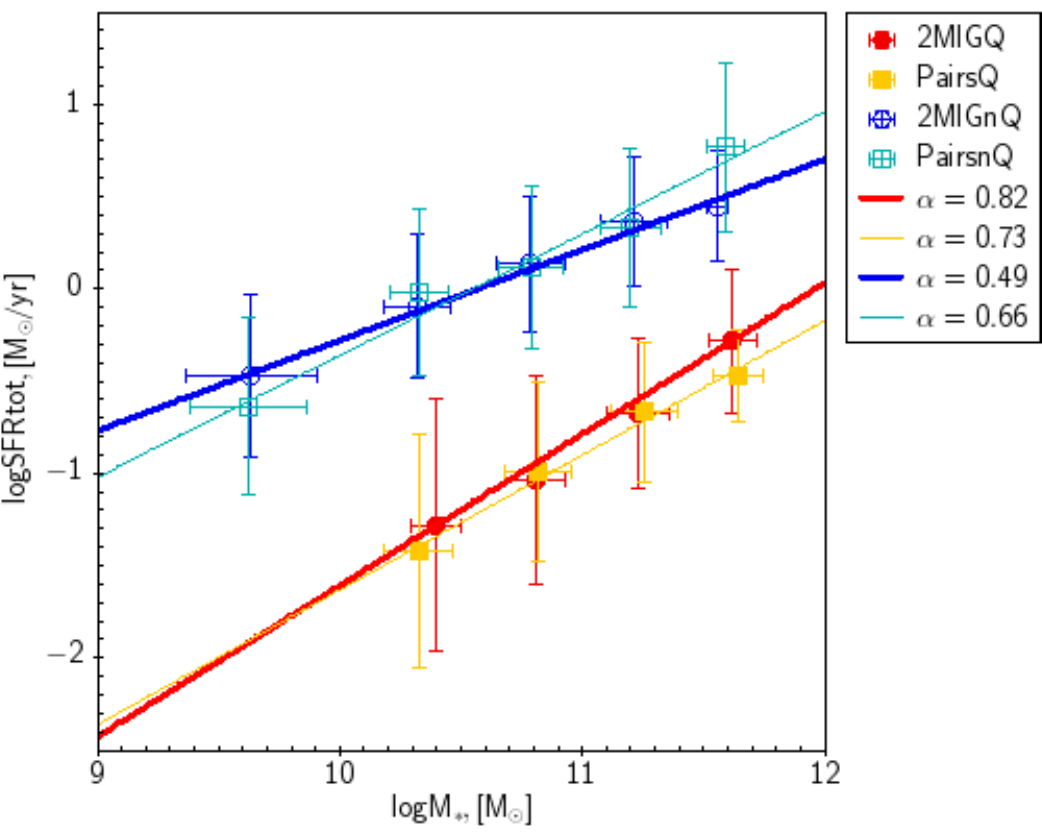} \\
\end{tabular}
\caption{Left: star formation rate (FUV) as a function of stellar mass for the 2MIG and paired galaxies.
Right: total (FUV+IR) star formation rate as a function of stellar mass. 
Different sub-samples (Q -- quenched and nQ -- non-quenched defined by FUV-K (AB) = 6.6 colour threshold) are marked with different colours.
The points present the mean values in the five stellar mass ranges (see text) while the error bars correspond to the 1 $\sigma$ standard deviation. 
The slopes of linear regression for each sub-sample are noted in the figures. The numbers
of galaxies in each sample are noted in Table 5.}
\label{9}
\end{figure*}

\begin{figure*}
\tabcolsep 0 pt
\begin{tabular}{cc}
\includegraphics[width=0.45\textwidth,natwidth=350,natheight=240]{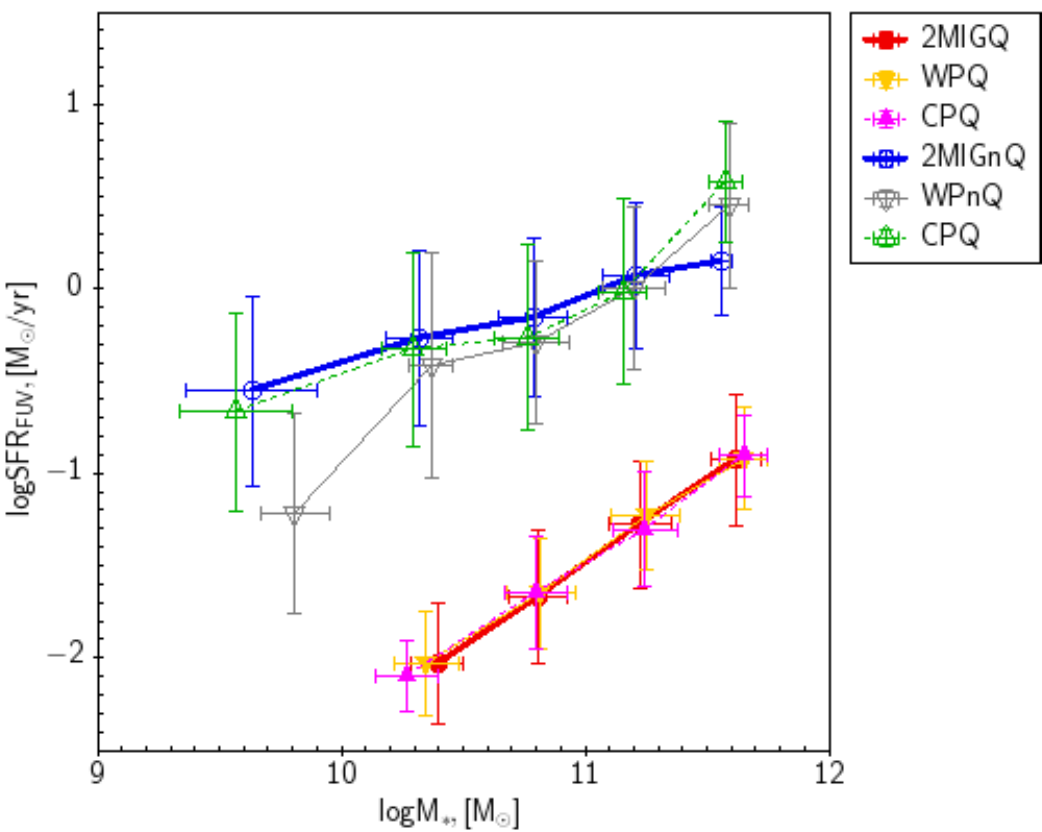} & 
\includegraphics[width=0.45\textwidth,natwidth=350,natheight=240]{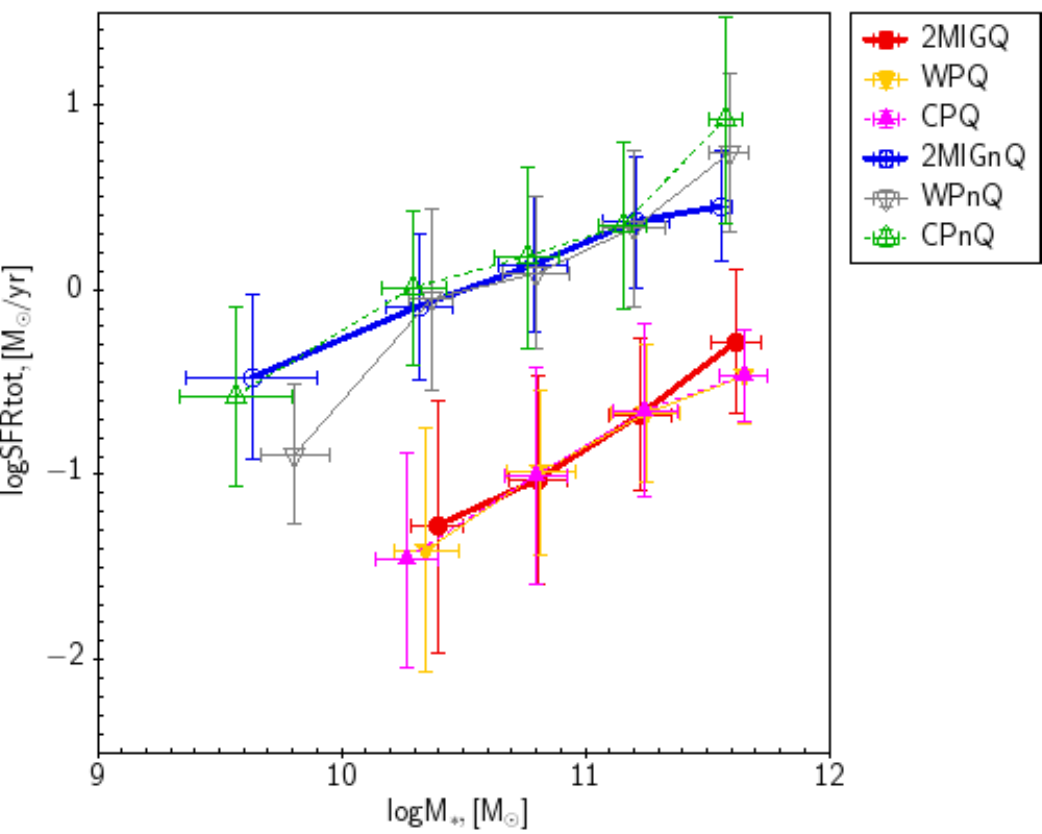} \\
\end{tabular}
\caption{Left: star formation rate (FUV) as a function of stellar mass for the 2MIG galaxies and galaxies from the wide and compact pairs (WP and CP, respectively).
Right: total (FUV+IR) star formation rate as a function of stellar mass. 
Different sub-samples (Q -- quenched and nQ -- non-quenched defined by FUV-K (AB) = 6.6 colour threshold) are marked with different colours.
The points present the mean values in the five stellar mass ranges (see text) while the error bars correspond to the 1 $\sigma$ standard deviation. 
The numbers
of galaxies in each sample are noted in Table 5.}
\label{10}
\end{figure*}

\begin{figure*}
\tabcolsep 0 pt
\begin{tabular}{cc}
\includegraphics[width=0.45\textwidth,natwidth=350,natheight=240]{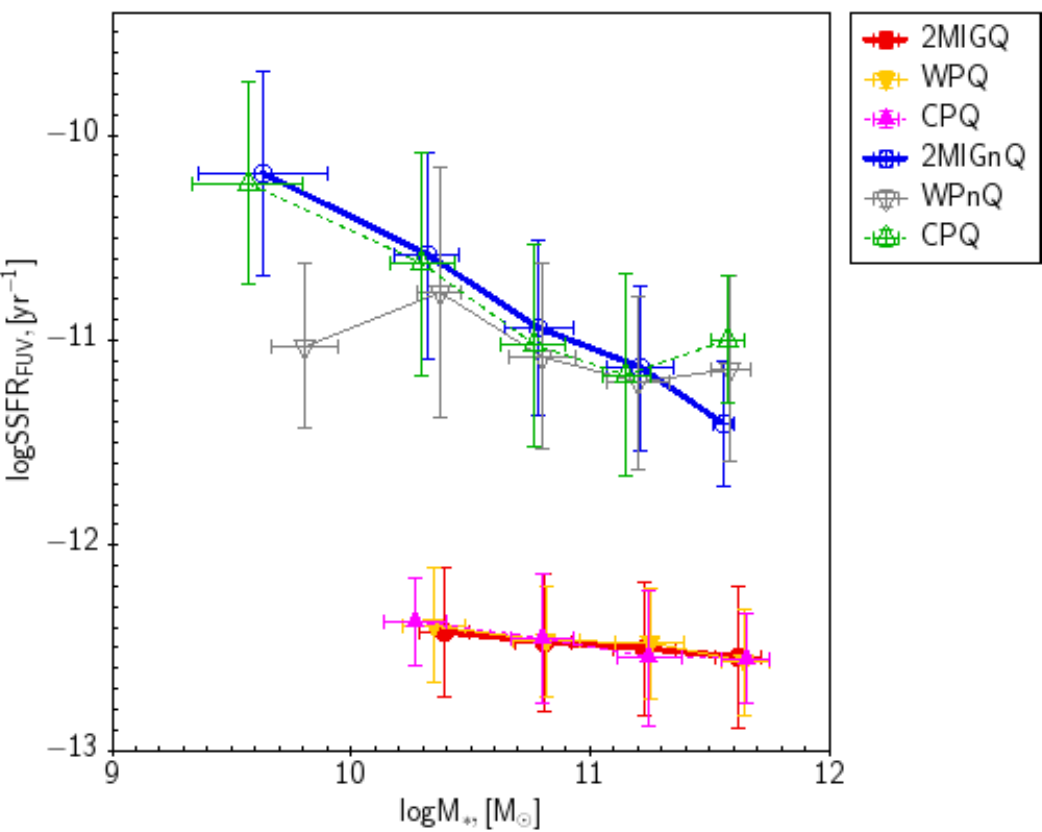} & 
\includegraphics[width=0.45\textwidth,natwidth=350,natheight=240]{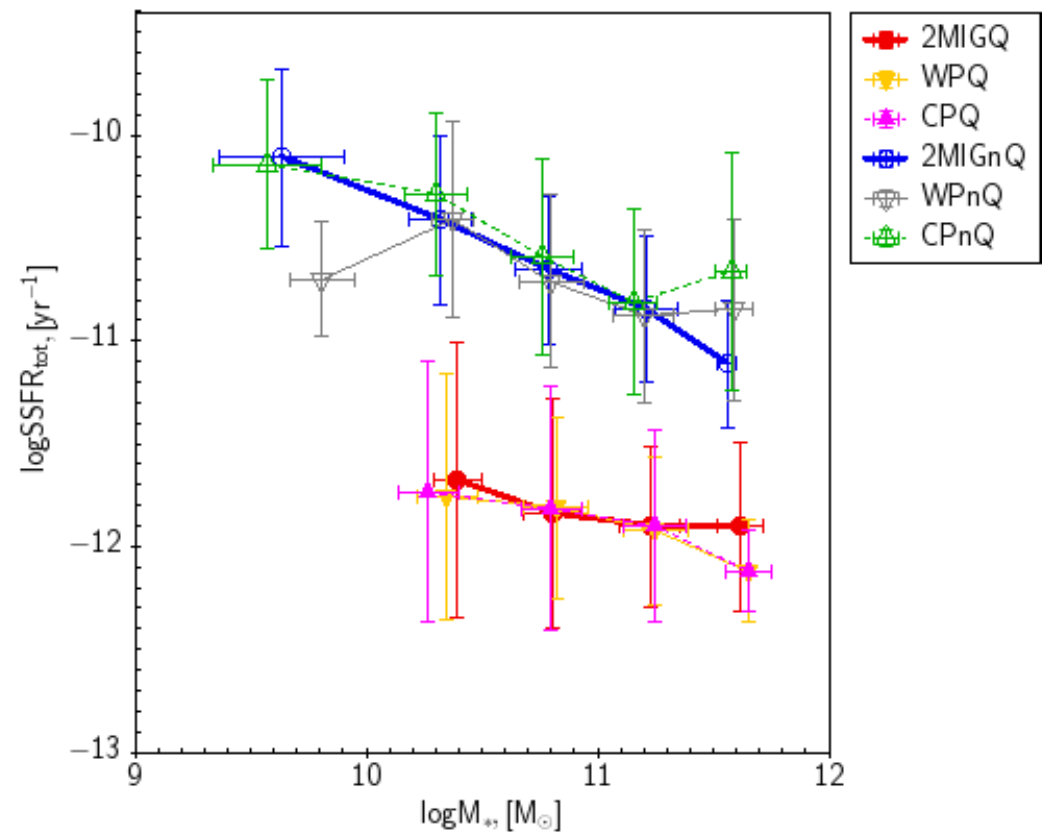} \\
\end{tabular}
\caption{Left: specific star formation rate (FUV) as a function of stellar mass for the 2MIG galaxies and galaxies from the wide and compact 
pairs (WP and CP, respectively).
Right: total (FUV+IR) specific star formation rate as a function of stellar mass. 
Different sub-samples (Q -- quenched and nQ -- non-quenched defined by FUV-K (AB) = 6.6 colour threshold) are marked with different colours.
The points present the mean values in the five stellar mass ranges (see text) while the error bars correspond to the 1 $\sigma$ standard deviation. The numbers
of galaxies in each sample are noted in Table 5.} 
\label{11}
\end{figure*}

\subsection{Morphology, mass, $SFR$ and $SSFR$ comparison}

In general, the sample of paired galaxies contains far more E type galaxies than the 2MIG sample (56\% vs. 31\%), 
see also discussion in Melnyk et al. (2014). We therefore observe a kind of morphology-density relation (Dressler 1980)
with the result that in higher populated regions the fraction of early-type galaxies is higher. A similar relation 
was previously noted by Karachentsev (1987) for paired vs. isolated galaxies and Vavilova et al. (2009) for galaxies in triplets in comparison 
with isolated ones. Therefore we see that the most effective quenching occured in 
high density regions. This was also noted in many other previous works, see for example Cucciati et al. (2006), Scoville et al. (2013), Tal et al. (2014).

Fig. 8 (left panel) presents the $FUV-K$ (AB) colour distribution for the 2MIG and paired galaxies, while the right panel represents 
the coincidence between ``colour'' types and visually defined types. The red line FUV-K (AB)=6.6 defines a chosen border between so called 
quenched (Q) and non-quenched (nQ) galaxies (see also Fig. 4). A detailed comparison between visual morphology and Q/nQ types are 
given in subsection 3.2. 

From the comparison of mean star formation characteristics of the 2MIG and paired 
galaxies (Tables 1 and 4, respectively) we can conclude that paired galaxies have lower values of $SFR$ and $SSFR$ for all sub-samples, 
excluding L types, where the star formation rates of paired galaxies are clearly higher. However, in spite of similar selection criteria 
described in the previous subsection, paired galaxies appear to be slightly more massive than isolated galaxies 
(this was reported also in Melnyk et al. 2014). For example, the mean log$M_*\pm$SD for L-type galaxies of Pairs and 2MIGs are 10.59$\pm$0.53 
and 10.27$\pm$0.69, respectively.  Therefore, in order to conduct an adequate comparison of $SFR$ and $SSFR$ between two samples, 
we calculated the corresponding values in the 
narrow ranges of stellar masses (9$<$log$M_*<$10, 10$<$log$M_*<$10.5, 10.5$<$log$M_*<$11, 11$<$log$M_*<$11.5 and 11.5$<$log$M_*<$12) for the Q
and nQ galaxy types, separately.

\begin{table}
\caption{The mean values and standard deviations of star formation and specific star formation rates of paired galaxies.
The morphological type division corresponds to E (-2$\leq$T$\leq$1), M (2$\leq$T$\leq$5) and L 
(6$\leq$T$\leq$10) types, Q for the quenched (FUV-K (AB)$>$6.6) and nQ for the non-quenched (FUV-K (AB)$<$6.6) galaxies.} 
\tabcolsep 1 pt
\begin{tabular}{lccccc} \hline
Sample  & N & log$SFR_{FUV}$ & log$SSFR_{FUV}$ & log$SFR_{tot}$ & log$SSFR_{tot}$\\
  & & [$M_{\odot}/yr$] & [yr$^{-1}$] & [$M_{\odot}/yr$] & [yr$^{-1}$] \\
   \hline
  All & 1482 &  -0.80$\pm$0.75 &  -11.86$\pm$0.81 &   -0.33$\pm$0.69 &  -11.39$\pm$0.75 \\
 E & 923 &   -1.15$\pm$0.59 &  -12.29$\pm$0.56 &   -0.61$\pm$0.60 &  -11.75$\pm$0.59 \\
 M & 505 &   -0.26$\pm$0.62 &  -11.23$\pm$0.61 &   -0.11$\pm$0.54 &  -10.85$\pm$0.63 \\
 L & 54 &   0.23$\pm$0.45 &  -10.37$\pm$0.53 &   0.46$\pm$0.48 &  -10.14$\pm$0.42 \\
 \hline
Q & 817 &   -1.32$\pm$0.45 &  -12.50$\pm$0.28 &   -0.75$\pm$0.50 &  -11.93$\pm$0.40 \\
nQ & 665 &   -0.15$\pm$0.52 &  -11.07$\pm$0.51 &   0.20$\pm$0.49 &  -10.71$\pm$0.47 \\
 \hline
\end{tabular}
\end{table}

\begin{table}
\caption{The Student's t-test probabilities compare the log$SFR_{FUV}$, log$SSFR_{FUV}$, log$SFR_{tot}$,
and log$SSFR_{tot}$ mean values for the nQ 2MIG and Pairs/CP/WP sub-samples in the lowest and highest mass ranges. 
These probabilities show that the 2MIG and Pairs sub-samples are drawn from the same parent set, having the same mean values.}
\tabcolsep 1 pt
\begin{tabular}{lccccc} 
\hline
\multicolumn{6}{c}{9$<$log$M_*$, [$M_{\odot}$]$<$10} \\				
Samp.1/2 & $N_1/N_2$  & log$SFR_{FUV}$ & log$SSFR_{FUV}$ & log$SFR_{tot}$  & log$SSFR_{tot}$ \\
 & & [$M_{\odot}/yr$] & [yr$^{-1}$] & [$M_{\odot}/yr$] & [yr$^{-1}$] \\
  \hline
2MIG/Pairs &	91/20 &	0.198 &	0.295 &	0.152 &	0.197 \\
2MIG/CP	& 91/16 & 0.602	& 0.627 & 0.738	& 0.773 \\
2MIG/WP	& 91/4 & 0.135 & 0.189 & 0.028 & 0.025 \\
  \hline
\multicolumn{6}{c}{11.5$<$log$M_*$, [$M_{\odot}$]$<$12} \\		
  \hline
2MIG/Pairs &	35/30 &	0.001 &	0.002	& 0.003	& 0.005 \\
2MIG/CP	& 35/4 & 0.105 & 0.237 & 0.103 & 0.268 \\
2MIG/WP	& 35/26	& 0.004	& 0.004	& 0.011	& 0.012 \\
  \hline
\end{tabular}
\end{table}

Fig. 9 presents the relation between the star formation rates of log$SFR_{FUV}$ on the left, log$SFR_{tot}$ on the right and the stellar mass
with the mean values with standard 1$\sigma$ deviations. For the main sequence galaxies we see the difference in
slopes $\alpha$ for the 2MIG and paired galaxies: the gradient for the former sample is smaller than the latter. 
For comparison, the fits for the Millenium simulations (0$<$z$<$0.3) and SDSS data from Elbaz et al. (2007) have slopes of $\alpha$=0.82 and 0.77, 
respectively, while the Boissier et al. (2010) model for z=0 fits with slope $\alpha$=0.65. 
Herewith the low mass nQ isolated galaxies have higher level of $SFR$ than paired galaxies while massive paired galaxies have somewhat higher $SFR$  than  isolated galaxies.  
In Fig. 10 we show the same values as in Fig. 9 for the 2MIGs but we split the sample of paired galaxies into the wide and compact pair members (WP and CP).
We see that the values of log$SFR_{FUV}$ and log$SFR_{tot}$ for nQ galaxies are 
usually higher for the CP members than for galaxies in the WP sample. This reveals a possible triggering effect due to interaction in the closer pairs. 

Table 5 shows the significance of the difference in the low and high mass ends for the nQ galaxies where visual differences in the Figs. 9,10
are prominent. The Student's t-test probabilities in the columns mean that the sub-samples of the 2MIGs and Pairs/WP/CP
drawn from the same parent sample and having the same mean values. The Student's t-test shows that the differences between $SFR$ and $SSFR$ in 
11.5$<$log$M_*<$12 mass range are more convincing than at 9$<$log$M_*<$10 masses.

Fig. 11 presents the specific star formation rate vs. stellar mass comparing Q and nQ types of the 2MIG, WP and CP samples in the 
stellar mass ranges similarly to Fig. 10. It can be seen
that the level of $SSFR$ in nQ isolated galaxies is somewhat higher than in paired galaxies,
supporting the previous results that galaxies located in low density regions have a higher level of specific star formation rate 
than those from denser regions (Rojas et al. 2004, Patiri et al. 2006, Elyiv et al. 2013). A difference between $SSFR$ of galaxies in cluster and field were 
found also by Haines et al. (2013) 
for higher redshift galaxies (0.15$<z<$0.2). The authors found that the starforming cluster galaxies have systematically 30\% lower $SSFR$ than field galaxies at all mass ranges. 
However, this difference is negligible between galaxies in the cluster outskirts and in the field. Our results are in nominal agreement with Haines's 
results in the low mass end. However, our findings suggest that the most massive paired galaxies 
11.5$<$log$M_*<$12  have a higher $SSFR$, triggering star formation. Note that our pairs may be located as in clusters as in the field.
The level of $SSFR$ for quenched galaxies is the same for the isolated and paired objects, showing that there is no environmental influence 
for these almost empty gas galaxies.  

Table 6 represents the slopes $\alpha$ of linear regressions for the 2MIG and paired galaxies. Schiminovich et al. (2007) and Salim et al. (2007) gave
the $\alpha$=-0.36 and -0.35 for the main sequence and -0.2 for the red sequence. Schawinski et al. (2014) noted about -0.1 $\div$ -0.2 slope for 
all SDSS galaxies. From our results it follows that the slope for the nQ 2MIGs is somewhat steeper than the slope for paired galaxies and  
noted values from Schiminovich et al. (2007) and Salim et al. (2007). Only two galaxies in our pairs have log$SSFR$ a 
little higher than -9.4.

Therefore we suggest that on the lower mass end, galaxies from denser environments have
an older stellar population which we can interpret as a sign of an environmental quenching. However, at higher masses, 
the positive influence of the environment, which triggers starformation, becomes remarkable. It follows from our findings that the main factor of evolutionary processes is
mainly defined by the mass-factor, though the environmental influence is notable. It is possible that we are unable to define the difference 
with better significance due to the smaller environmental effect in comparison with mass-influence. In general, we are in agreement 
with the results obtained in previous works by other authors, see for example Peng et al. (2010, 2012) and Tal et al. (2014).

\subsection{AGN impact}

Karachentseva et al. (2014) studied the $SFR$ of the Markarian galaxies, which are active galaxies with 
different signs of activity (Markarian et al. 1989, Mazzarella \& Balzano 1986, Petrosian et al. 2007): quasars, Seyfert galaxies, 
Wolf-Rayet, Starburst and HII galaxies. The authors found that $SFR$ in Markarian galaxies is a little higher than in isolated galaxies, 
though the $SSFR$ does not exceed a limit of $\sim$dex(-9.4) [-yr$^{-1}$], also noted in subsection 3.4 of this paper. 
In the previous section we found some shortage/excess of $SFR$ (and also $SSFR$) according to 
environmental events for low mass/high mass galaxies. Here we would like to examine how it is related to the AGN activity. 
For this we check and compare the AGN presence in isolated and paired galaxies.

\begin{table}
\caption{The slope parameter in the log$SSFR$, [yr$^{-1}$] vs. log$M_{*}$, [$M_{\odot}$] relations.} 
\tabcolsep 4 pt
\begin{tabular}{lccccc} \hline
Sample & N & $\alpha$, (log$SSFR_{FUV}$)& $\alpha$, (log$SSFR_{tot}$) \\
  \hline
  \multicolumn{4}{c}{2MIG}  \\
Q & 341 & -0.09  & -0.18 \\
nQ & 1275 & -0.63  & -0.50 \\

  \hline
\multicolumn{4}{c}{Pairs}  \\
Q & 817 & -0.13  & -0.27 \\
nQ & 665 & -0.41  & -0.34 \\
 \hline
\end{tabular}
\end{table}

Since we have the WISE magnitudes for almost all of our galaxies, 
we define the AGN fraction according to W2-W1 vs. W3-W2 colour diagrams, which is 
effective in selection of bright IR AGN (see for example, Jarrett et al. 2011), the 
majority of these sources should be also bright in X-rays (Mateos et al. 2012). Fig. 12 presents the WISE 
colour-colour plots for galaxies from the 2MIG (left panel) and WP/CP samples (right panel). Since near 1/2 of our sample are also SDSS targets, we
marked with different colour symbols SDSS-selected AGN type 1 and 2 on the left 
panel of Fig. 12. The AGN classification was carried out by Coziol et al. (2011), Pulatova et al. (2015) and a few additional AGN sources were found in the NED. 
Of course this spectroscopic classification is not homogeneous, rather only indicative. We marked with black dots any galaxies which do
not contain AGN or where we have no information about AGN presence. In Fig. 12 we also marked with lines different criteria for the selection of 
the brightest IR AGN from Mateos et al. (2012), Stern et al. (2012) and Hwang et al. (2012).  

In Fig. 13 we show the fractions of AGN according to the weakest criterion by Hwang et al. (2012; W1-W2$>$0.44) for the 2MIG and full paired sample 
in different mass intervals (on the left) and also separately for Q and nQ sub-samples (on the right). These uncertainties correspond to 
95\% confidence level ($2\sigma$) calculated with the Binomial test. We see that the fraction of AGN in 
paired samples is almost always higher than in an isolated galaxy sample. However, the advantage is not crucial to conclude that the AGN phenomenon is 
necessarily connected with environmental density. We also did not find any increase in AGN fraction in most tight pairs as
shown by Ellison et al. (2011) though we do not have a large statistical sample on pairs with R$<$20 kpc. 

Here we are in agreement with findings by Coziol et al. (2011), Sabater et al. (2012), 
Hernandez-Ibarra et al. (2013) and Pulatova et al. (2015) who studied the 
nuclear activity in a few samples of isolated galaxies and showed that secular evolution is a main mechanism for AGN triggering, at 
least in the Local Universe. Here we have to make a reservation that in our sample of the local galaxies we do not have bright X-ray AGN which evolution may 
be explained by the merging events (see for example Santini et al. 2012).

\begin{figure*}
\tabcolsep 0 pt
\begin{tabular}{cc}
\includegraphics[width=0.55\textwidth,natwidth=400,natheight=230]{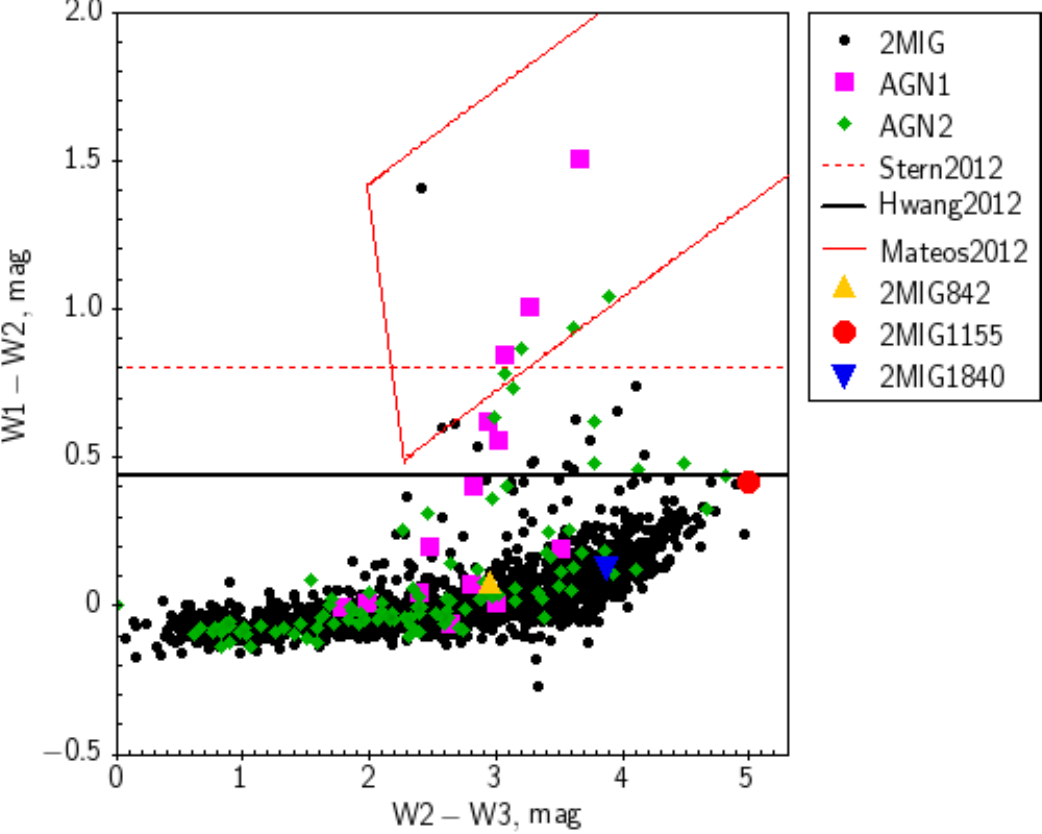} &
\includegraphics[width=0.55\textwidth,natwidth=400,natheight=230]{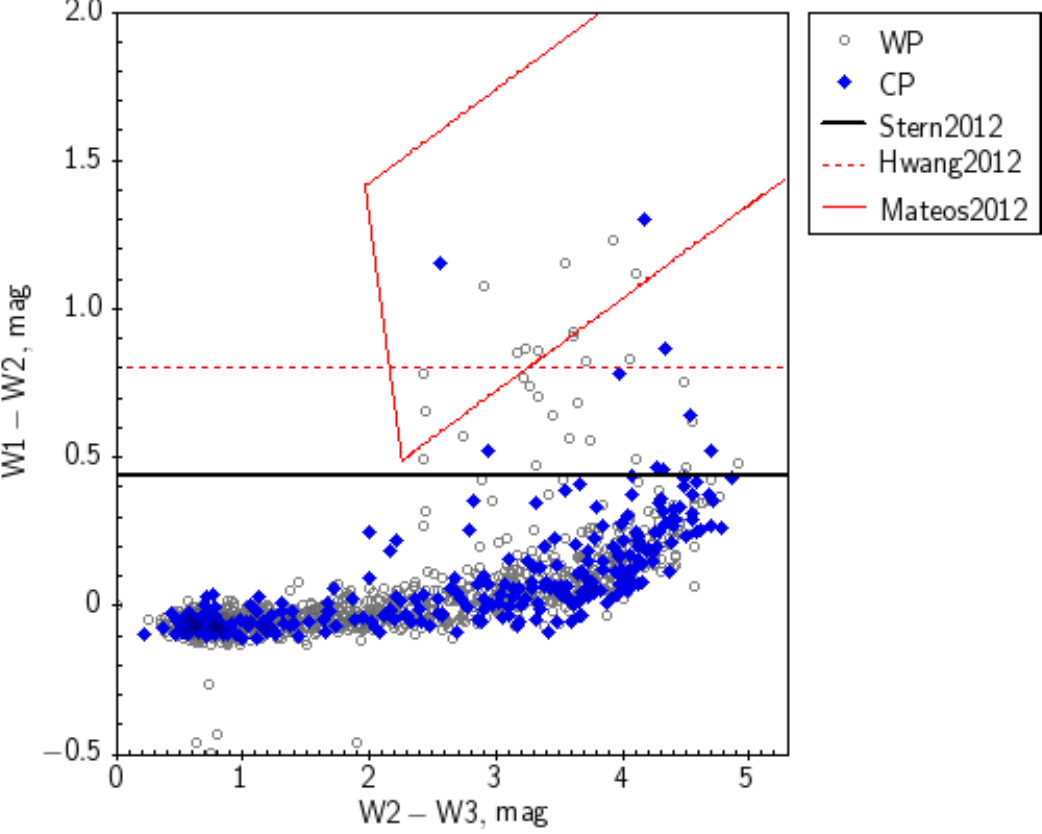} \\
\end{tabular}
\caption{Left: WISE colour-colour plot for 2MIG galaxies. Right: WISE colour-colour plot for paired galaxies. Red line selects the bright AGN 
according to criteria by Mateos et al. (2012), dashed line shows the selection according to Stern et al. (2012) and the solid line 
represents the Hwang et al.(2012) selection adopted in this work: W1-W2$>$0.44 mag. Three galaxies with the highest $SFR_{IR}/SFR_{FUV}$ ratios 
are shown with special symbols, according to Jarrett et al. (2011) diagnostic diagram of their IR emission attributes due to starbursts.}
\label{12}
\end{figure*}

\begin{figure*}
\tabcolsep 10 pt
\begin{tabular}{cc}
\includegraphics[width=0.35\textwidth,natwidth=450,natheight=380]{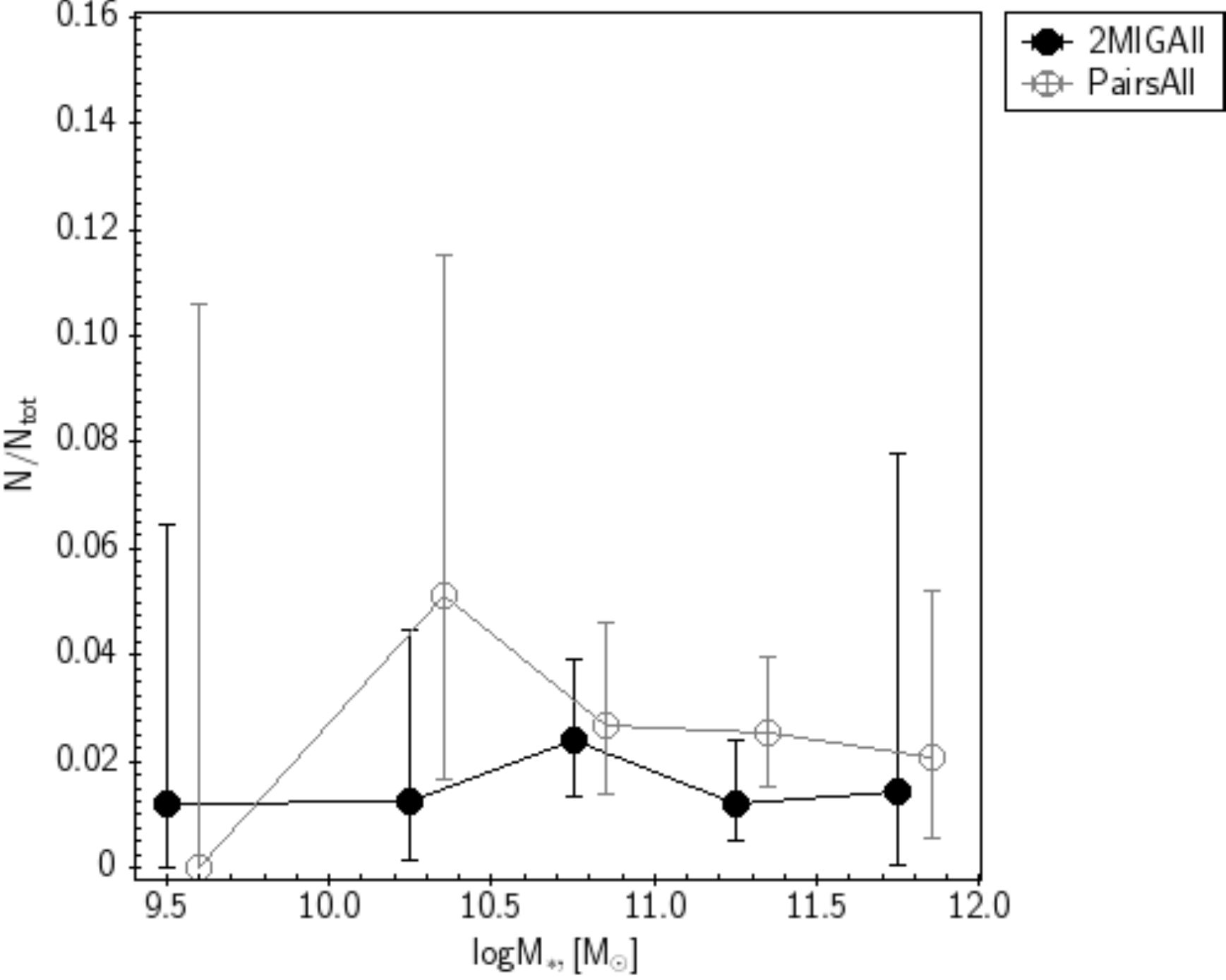} &
\includegraphics[width=0.35\textwidth,natwidth=450,natheight=380]{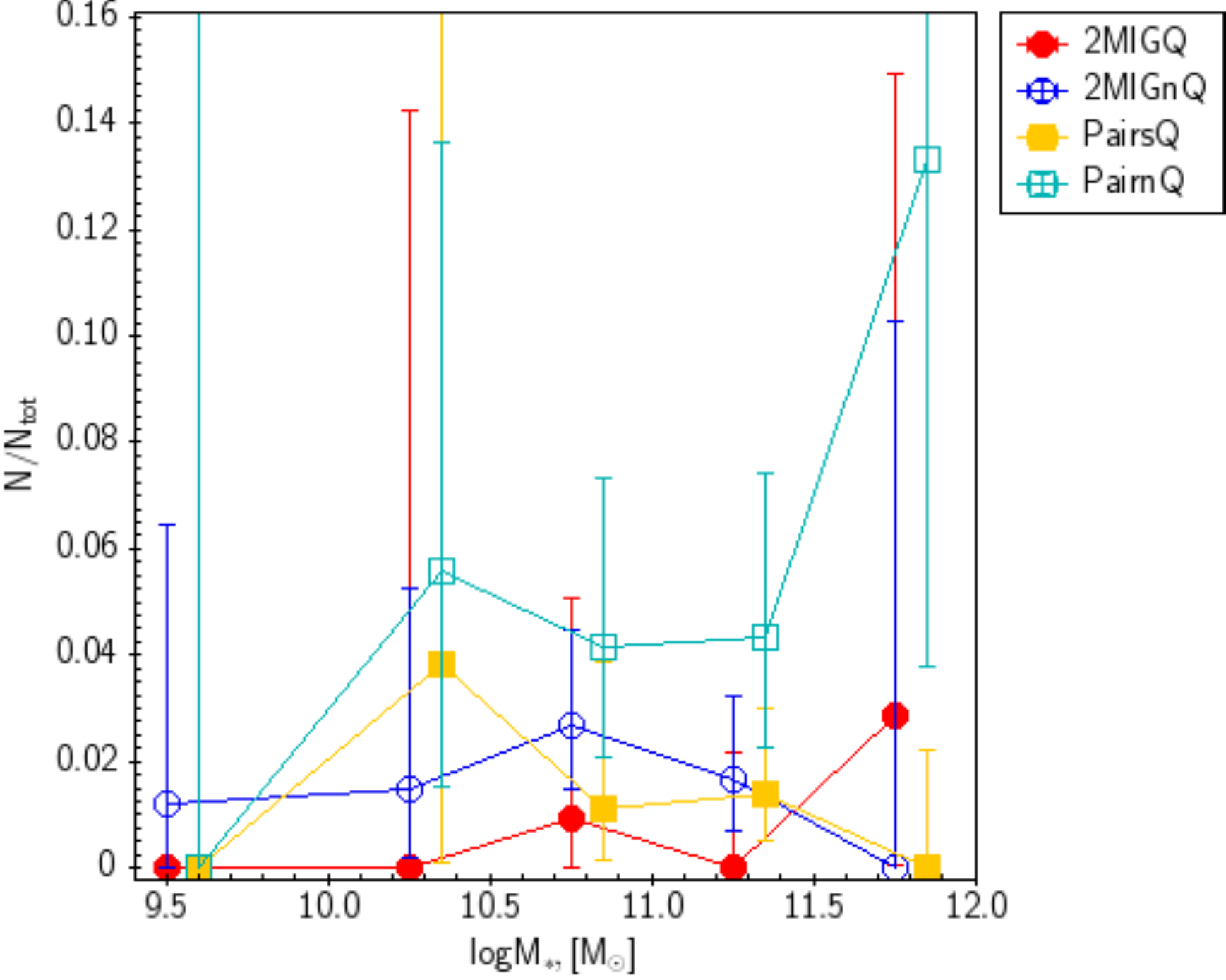} \\
\end{tabular}
\caption{Left: The fraction of AGN selected by the W1-W2$>$0.44 colour criterion (Hwang et al. 2012) for 
the 2MIG and Pairs samples in different mass ranges. The uncertainties correspond to 2$\sigma$ standard deviation. Right: 
The fraction of AGN as in left panel but for Q and nQ types separately. }
\label{13}
\end{figure*}

\section{Conclusions}

In the paper we have considered the star formation properties of 1616 isolated galaxies from the 2MIG catalogue 
(Karachentseva et al. 2010) with FUV magnitudes. The star formation rates ($SFR$) and specific star formation rates ($SSFR$) of 2MIG galaxies 
were compared with corresponding properties of isolated galaxies from the LOG catalogue (Karachentsev et al. 2011) and 
with paired galaxies (WP -- wide pairs and CP -- the most compact pairs with velocity difference less than 150 km/s and a projected
distance less than 50 kpc). We summarize our main findings below:

1. Different selection algorithms applied to different primordial samples define different populations of isolated galaxies. The population of the
LOG catalogue, selected from the Local Supercluster non-clustered galaxies, 
mostly consists of spiral and late type galaxies. These are up to two orders of magnitude
less massive than the 2MIG galaxies of corresponding morphology.
The 2MIG population, in сontrast,  was selected from the 2MASX (Jarrett et al. 2000) within quite bright near-infrared galaxies ($K_s<$12 mag and 
angular diameters $a_{Ks}\geq $ 30$''$) consisting of normal mass galaxies, predominantly of early spirals. Late type galaxies of both samples
have almost the same amount of gas at present epochs though the gas fraction in spiral and early type galaxies of 
the 2MIG sample is much smaller and will be consumed over less time than $T_0$. In general, the $SFR$ vs. $M_*$ and $SSFR$ vs. $M_*$ 
relations of the LOG sample complete 2MIG corresponding relation to lower masses.
 
2. The $SSFR$ upper limit in all considered galaxy samples (i.e. isolated and paired) 
of the Local Universe (at least at z$<$0.06) does not exceed the value of $\sim$dex(-9.4). This is probably common for galaxies
of differing activity located in different types of environments (Karachentsev et al. 2013, Karachentsev and Kaisina 2013, 
Karachentseva et al. 2014).

3. The fractions of quenched (Q, with FUV-K (AB) $>$ 6.6) galaxies of 10.5$<$log$M_*<$11.5 masses are nearly twice as high
in the paired galaxy sample than among the 2MIG isolated galaxies. This suggests that quenching proved more effective in higher density regions 
in the past (Cucciati et al. 2006, Scoville et al. 2013, Tal et al. 2014). From the behaviour of $SFR$ vs. $M_*$ relations we concluded 
that the main factor of evolutionary processes is defined by the galaxy mass. However, the environmental influence is notable: 
the 2MIG's and pair's main sequence (non-Quenched with FUV-K (AB) $<$ 6.6) galaxies have different gradients: 
the relation for the former sample is smaller than the latter, i.e. the paired galaxies have somewhat lower $SFR$ and $SSFR$ at low masses and 
higher $SFR$ and $SSFR$ for the massive galaxies with log$M_*>$11.5. The Student's t-test showed that the difference is significant only for the high-mass end. Therefore we suggest, 
that the environment triggers star formation in the highest mass galaxies. Moreover, the values of $SFR$ and
$SSFR$ are higher in general for galaxies in the most closely spaced pairs suggesting the outgoing interactions.

4. The existence of a significant fraction of quenched isolated galaxies (21\%) reveals that
the main quenching mechanism probably does not depend on the environmental density. The similarity of the mean values of the $SFR$ and gradients 
in $SFR$ vs. $M_*$ (also in $SSFR$ vs. $M_*$) relations for quenched 2MIG and paired galaxies also support the assumption that the 
dominant quenching mechanism is similar for galaxies located in all type of environments and depend mainly upon the galaxy mass 
(possibly the main driving mechanism is AGN feedback Croton \& Farrar 2008, see also discussion in Peng et al. 2010 and Schawinski et al. 2014).  

5.  We found that the fraction of AGN in paired samples is a little higher (but not significantly in the statistical sense) than in our isolated galaxy sample. We 
suggest that AGN phenomenon is not necessarily connected with environmental density and, most probably, defined by secular galaxy evolution
confirming the previous results by Coziol et al. (2011), Sabater et al. (2012), Hernandez-Ibarra et al. (2013), Karachentseva et al. (2014) and Pulatova et al. (2015).

\section*{Acknowledgments}
We are thankful to Samuel Boissier for his useful comments and suggestions. We are also grateful to Andrew Whimster for the language corrections.
This research has made use of the NASA/IPAC Extragalactic Database (NED) which is operated by the Jet Propulsion Laboratory, 
California Institute of Technology, under contract with the National Aeronautics and Space Administration. We acknowledge the usage 
of the HyperLeda database (http://leda.univ-lyon1.fr). In our work we have also used SDSS-III data, funding for which has been provided by 
the Alfred P. Sloan Foundation, the Participating Institutions, the National Science Foundation, and the U.S. Department of Energy Office of Science. 
The SDSS-III web site is http://www.sdss3.org/. This work was supported by a grant from the Russian Foundation for Basic 
Researches, 13-02-90407 Ukr-f-a.

\label{lastpage}

\end{document}